\documentclass[twocolumn,aps,prb,superscriptaddress,floatfix,noeprint]{revtex4-1}

\usepackage{graphicx,tabularx}
\usepackage{amsmath,amsfonts,amssymb}
\usepackage{mathtools}
\usepackage{bm,dsfont}
\usepackage[colorlinks=true, citecolor=blue, linkcolor=red, urlcolor=blue]{hyperref}
\usepackage[all]{hypcap}
\usepackage{xcolor}
\usepackage[caption=false]{subfig}

\graphicspath{{figures/}}

\newcolumntype{s}{>{\hsize=.1\hsize}X}
\newcolumntype{b}{>{\hsize=.45\hsize\centering\arraybackslash}X}

\begin{document}

\title{Network model and four-terminal transport in minimally twisted bilayer graphene}
\author{Christophe De Beule}
\affiliation{Institute for Mathematical Physics, TU Braunschweig, 38106 Braunschweig, Germany}
\affiliation{Department of Physics and Materials Science, University of Luxembourg, L-1511 Luxembourg, Luxembourg}
\author{Fernando Dominguez}
\affiliation{Institute for Mathematical Physics, TU Braunschweig, 38106 Braunschweig, Germany}
\author{Patrik Recher}
\affiliation{Institute for Mathematical Physics, TU Braunschweig, 38106 Braunschweig, Germany}
\affiliation{Laboratory for Emerging Nanometrology, 38106 Braunschweig, Germany}
\date{\today}

\begin{abstract}
We construct a two-channel scattering model for the triangular network of valley Hall states in interlayer-biased minimally twisted bilayer graphene from symmetry arguments and investigate electronic transport in a four-terminal setup. In the absence of forward scattering, a single  phenomenological parameter tunes the network between a triplet of chiral zigzag modes and pseudo-Landau levels. Moreover, the chiral zigzag modes give rise to robust Aharonov-Bohm resonances in the longitudinal conductance in the presence of a perpendicular magnetic field or an in-plane electric field. Interestingly, we find that when both a magnetic field and an in-plane electric field are applied, the resonances of different zigzag branches split depending on their propagation direction relative to the in-plane electric field. We further demonstrate that while the Hall response vanishes in the chiral zigzag regime, a finite Hall response is obtained without destroying the Aharonov-Bohm resonances in the longitudinal response, by weakly coupling different zigzag branches, which also gives rise to Hofstadter physics at accessible magnetic fields.
\end{abstract}

\maketitle

\section{Introduction}

In recent years, twisted bilayer graphene \cite{LopesDosSantos2007,Li2010} has attracted great interest from the condensed-matter community. In this system, two graphene layers are rotationally mismatched, giving rise to a moir\'e pattern in the layer stacking. In a large part, this interest was motivated by the experimental discovery of superconductivity and correlated phases \cite{TramblyDeLaissardiere2010,SuarezMorell2010,Bistritzer2010,Kim2017a,Cao2018a,Cao2018,Yankowitz2019,Sharpe2019,Kerelsky2019,Choi2019,Cao2019} at the ``magic" twist angle ($\theta\sim 1^\circ$). 
For tiny twist angles ($\theta\sim 0.1^\circ$) also known as minimally twisted bilayer graphene (mTBG), the moir\'e pattern grows so large that it becomes energetically favorable to shrink hexagonal stacking regions where locally all carbon atoms are eclipsing (AA stacking) in favor of expanding the Bernal-stacked regions (AB or BA stacking) where only atoms of different sublattices overlap, at the expense of intralayer strain \cite{Nam2017,Walet2019}. The relaxed structure is then given by a triangular tiling of alternating AB and BA regions, see Fig.~\ref{fig:lattice}(a), whose vertices correspond to AA regions. The latter act as topological defects, giving rise to three AB/BA domain walls intersecting at each node \cite{Alden2013}. Upon application of an interlayer bias, e.g., due to an electric field perpendicular to the layers, a local gap is opened in the Bernal-stacked regions, while the AA regions remain metallic. When the Fermi level lies in the local gap, one can thus think of mTBG as a triangular lattice of quantum dots. Moreover, the AA regions are coupled via the AB/BA domain walls which support two chiral modes per valley and spin, and which are helical in the sense that states in opposite valleys counterpropagate, see Fig.~\ref{fig:lattice}(b). The existence of these chiral modes can be understood from the change in valley Chern number $\Delta N_K = -\Delta N_{K'} = \pm 2$ across an AB/BA or BA/AB domain wall, respectively. Here, the valley Chern number is defined locally in real space and momentum space, and therefore it is not necessarily quantized \cite{Zhang2013,San-Jose2013}. Nevertheless, the change in valley Chern number across a domain wall is quantized \cite{Zhang2013}. From the bulk-boundary correspondence \cite{Hasan2010}, we require two chiral modes per valley and spin that propagate along the domain walls, where the propagation direction is opposite for opposite valleys. 
The low-energy physics is thus captured by a triangular network of chiral modes where the AA regions act as scattering centers \cite{San-Jose2013,Efimkin2018,Ramires2018,Huang2018,Sunku2018,Rickhaus2018,Xu2019,Yao2020,Verbakel2021}, which is illustrated in Fig.~\ref{fig:lattice}(c). Assuming the valleys are decoupled, we have (for each spin) two triangular networks of opposite orientation, related by time-reversal symmetry. For a given valley, this system is reminiscent of a triangular \cite{Mkhitaryan2009} Chalker-Coddington oriented network \cite{Chalker1988,Kramer2005}.

Recently, robust Aharonov-Bohm (A-B) oscillations attributed to the network were observed in transport experiments in interlayer-biased mTBG \cite{Xu2019}. The A-B oscillations were observed on top of a constant plateau and persisted at finite temperatures below the local gap. In addition, oscillations of the Hall resistivity were also observed. The latter are not robust against temperature and can be qualitatively understood in terms of network bands of alternating electron and hole character. On the theory side, microscopic calculations demonstrated that the network hosts so-called one-dimensional (1D) chiral zigzag (ZZ) modes \cite{Tsim2020,DeBeule2020a}. In this regime, the network (for a given valley and spin) effectively splits in three independent families of 1D chiral modes, where each family consists of modes propagating in parallel, giving rise to a nested Fermi surface between valleys \cite{Fleischmann2020}. Moreover, in the presence of a magnetic field perpendicular to the graphene layers, scattering between parallel ZZ modes gives rise to robust A-B oscillations in the longitudinal conductance \cite{DeBeule2020a}.  However, given the 1D nature of these chiral ZZ modes, one expects a vanishing Hall response, at odds with experiment. In this paper, we show that this dilemma is resolved when different ZZ branches are weakly coupled, in which case the A-B oscillations persist and the Hall response can be nonzero.

This paper is organized as follows. In Sec.\ \ref{sec:model}, we introduce the network model. We take a phenomenological approach and obtain the general constraints on the $S$ matrix from symmetry arguments. First, we consider the case without forward scattering at the nodes. In this case, the $S$ matrix depends on a single parameter, which tunes the network in mTBG between a triplet of one-dimensional chiral zigzag modes \cite{Fleischmann2020,Tsim2020,DeBeule2020a} and pseudo-Landau levels \cite{Ramires2018}. We then allow for forward scattering and discuss different processes between zigzag modes. Coupling between parallel zigzag channels only warps the Fermi surface, while a gap is generically opened due to scattering between zigzag modes that propagate in different directions. In Sec.\ \ref{sec:transport}, we investigate electronic transport in the chiral zigzag regime in a four-terminal setup in the presence of a magnetic field perpendicular to the layers, as well as when a uniform in-plane electric field is applied to the system. Here, forward scattering gives rise to Aharonov-Bohm oscillations in the longitudinal conductance, and we show that resonances of different zigzag branches are split by the in-plane electric field. This gives rise to magneto-electric Aharonov-Bohm oscillations. Similar such oscillations were recently observed in mTBG by STM measurements \cite{Liu2021}. We also discuss magnetotransport in the percolating regime, i.e., when different zigzag branches are coupled, giving rise to Hofstadter physics at experimentally accessible magnetic fields. Finally, we present our conclusions in Sec.\ \ref{sec:conclusions}.

\section{Network model} \label{sec:model}

\begin{figure}
\centering
\includegraphics[width=\linewidth]{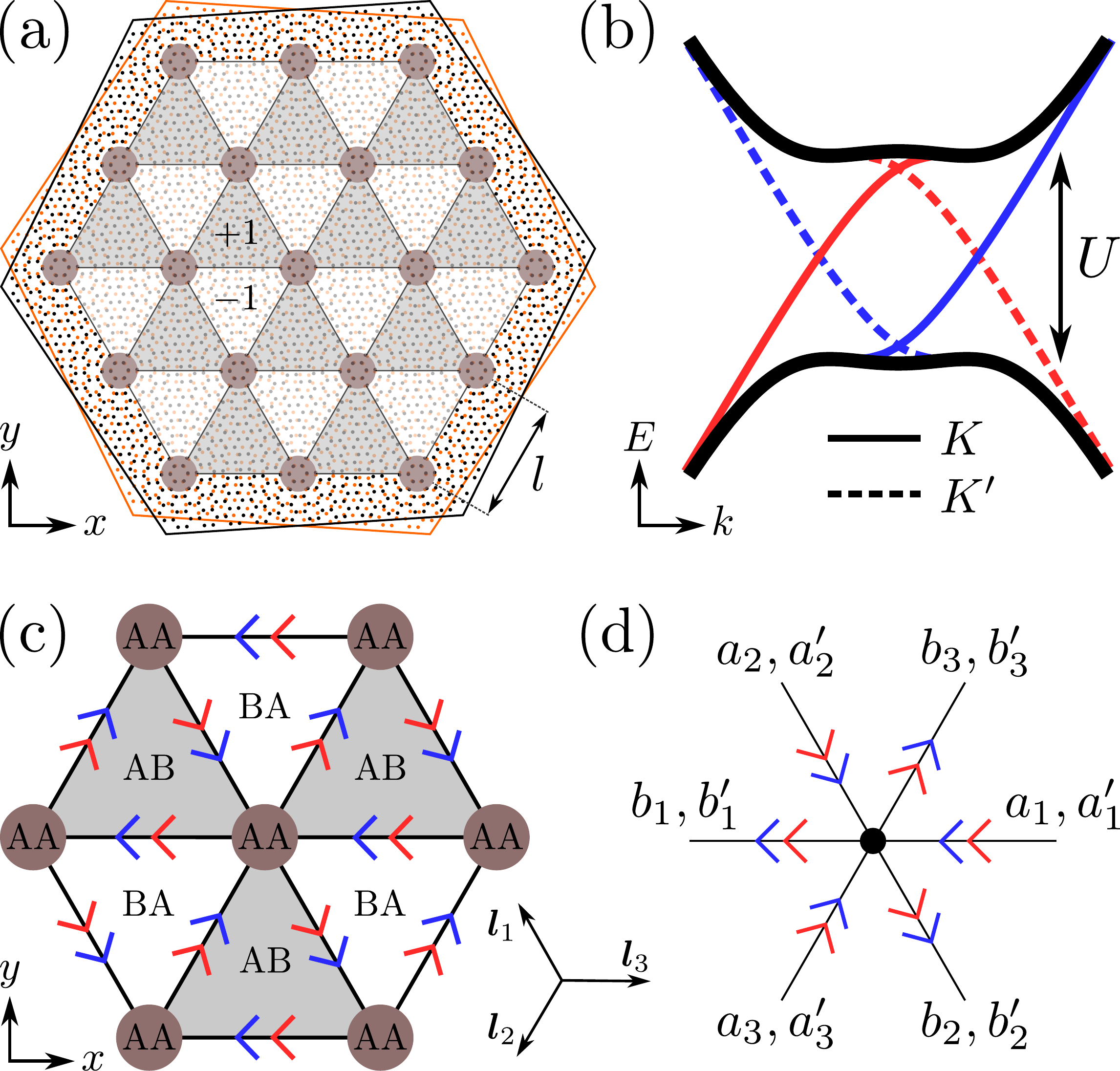}
\caption{(a) Illustrative example of the TBG lattice for $\theta=7.34^\circ$ where atoms of top and bottom layers are shown as black and orange dots. Note that this is not a minimal twist angle. The AB, BA, and AA stacking domains are shown as gray, white, and brown regions and $N_K=-N_{K'}\approx\pm1$ is the local valley Chern number in the presence of an interlayer bias $U$. (b) Dispersion along domain walls between two semi-infinite AB and BA regions, calculated with the four-band continuum model for $U/\gamma_\perp=0.1$, showing two chiral modes (blue and red lines) per valley and spin. (c) Network of valley Hall states (for a single valley and spin) where domain walls and AA regions correspond to links and scattering nodes of the network, respectively. (d) Unit cell of the network.}
\label{fig:lattice}
\end{figure}

The network in mTBG consists of scattering nodes (AA stacking regions) that form a triangular lattice and links between these nodes (AB/BA domain walls), as illustrated in Fig.~\ref{fig:lattice}(a). Each link hosts two valley Hall channels (for a given valley and spin) [Fig.~\ref{fig:lattice}(b)] such that each scattering node has six incoming and six outgoing modes, as shown in Fig.~\ref{fig:lattice}(c). 
We label the nodes by a pair of indices ($m,n$) with position $m \bm l_1 + n \bm l_2$, where $\bm l_{1,2} = l (-1/2,\pm \sqrt{3}/2)$ are moir\'e lattice vectors with $l = a/2\sin(\theta/2) \approx 14 ( \theta^\circ )^{-1}$~nm the moir\'e lattice constant and $a \approx 0.25$~nm the graphene lattice constant of graphene. The scattering amplitudes of the incoming modes at node $(m,n)$ are then written as
\begin{equation}
a_{mn} = \left( a_{1mn}, a_{1mn}', a_{2mn}, a_{2mn}', a_{3mn}, a_{3mn}' \right)^t,
\end{equation}
where the numbers $1$, $2$, and $3$ indicate the link as defined in Fig.~\ref{fig:lattice}(d) and the primed amplitudes  belong to the second channel (not be confused with the valley). The outgoing amplitudes are defined in the same way, and denoted as $b_{mn}$. They are related to the incoming modes via the $S$ matrix,
\begin{equation} \label{eq:Smat}
b_{mn} = \mathcal S a_{mn},
\end{equation} 
which characterizes the nodes of the network. Current conservation at a node is expressed as $\mathcal S^\dag\mathcal S = \mathds 1_6$.

In this work, we do not consider intervalley scattering as the moir\'e pattern varies slowly on the interatomic scale for small twists. Hence, we consider two independent networks, one for each valley, with opposite orientations as the valleys are related by time-reversal symmetry. 

\begin{figure*}
\centering
\includegraphics[width=\linewidth]{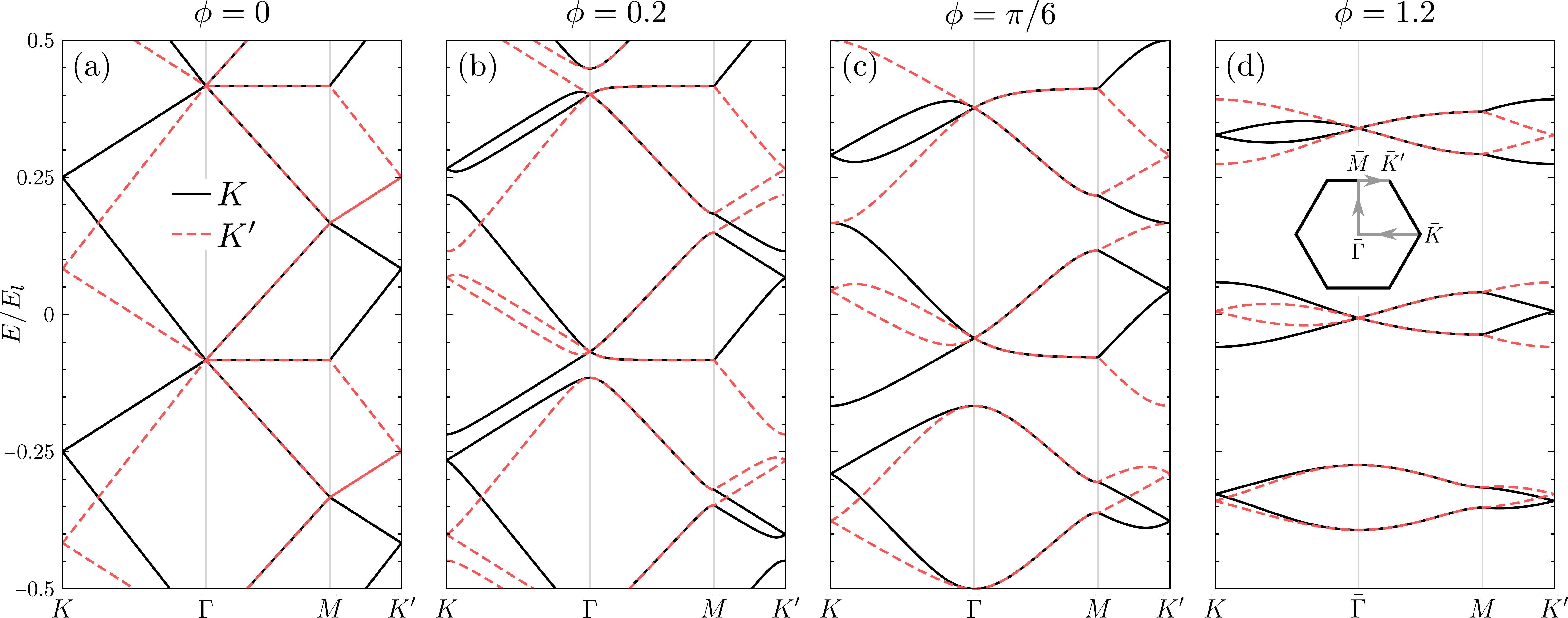} 
\caption{Network bands in the absence of forward scattering along high-symmetry lines [see inset in (d)] for the $K$ valley (solid black) and $K'$ valley (dashed red). Here, we have shifted the energy by an overall constant $E_l/12$. (a) Chiral zigzag regime. (b) Coupled zigzag modes. (c) Indirect gap opening. (d) Regime close to flatbands showing graphene-like bands.}
\label{fig:bands1}
\end{figure*}

\subsection{$S$ matrix}

To find the $S$ matrix, we take a phenomenological approach, where we constrain its form with unitarity and the symmetries of mTBG in the presence of an interlayer bias \cite{Po2018,Zou2018}. 
The symmetries of interlayer-biased TBG are given by $C_3$ and $C_2$ rotations about the $z$ axis with respect to the center of an AA region, and time-reversal symmetry $T$. While $C_3$ symmetry conserves the valley, both $C_2$ and $T$ flip the valley index. Within one valley, the symmetries are therefore given by $C_3$ and $C_2T$. 
The former corresponds to a cyclic permutation of the incoming amplitudes $a_1 \rightarrow a_2 \rightarrow a_3 \rightarrow a_1$, see 
Fig.~\ref{fig:lattice}(d). It follows that the $S$ matrix can be written as
\begin{equation} \label{eq:C3}
\mathcal S = \begin{pmatrix} s_f & s_l & s_r \\ s_r & s_f & s_l \\ s_l & s_r & s_f \end{pmatrix},
\end{equation}
where $s_f$, $s_r$, and $s_l$ contain the amplitudes for forward scattering and right and left deflections by $120^\circ$, subject to $\mathcal S^\dag \mathcal S = \mathds 1_6$. Furthermore, when additionally $C_2$ or $T$ is conserved, it can be shown (see App.\ \ref{app:model}) that the amplitudes of different valleys are related as follows: 
\begin{alignat}{3}
& C_2: \qquad && s_{r(l)} = s_{r(l)}', \quad && s_f = s_f', \\
& T: \qquad && s_{r(l)} = (s_{l(r)}')^t, \quad && s_f = (s_f')^t,
\end{alignat}
such that when the combined symmetry $C_2T$ is conserved, we have for a given valley
\begin{equation}
C_2 T: \qquad s_r =  (s_l)^t, \quad s_f = (s_f)^t.
\end{equation}
One can show that these conditions reduce the number of free parameters of the $S$ matrix to six real parameters (not including a global phase). We demonstrate this explicitly in App.\ \ref{app:model}, where we construct the most general $S$ matrix using a scattering basis that transforms properly under $C_3$ and $C_2$. 

\subsection{Network bands}

The incoming modes of a given node are related to outgoing modes of neighboring nodes,
\begin{equation}
\begin{aligned}
a_{mn} & = e^{i2\pi E/E_l} \left( b_{1m-1n-1}, b_{1m-1n-1}', \right. \\
& \hspace{1.85cm} \left. b_{2m+1n}, b_{2m+1n}', b_{3mn+1}, b_{3mn+1}' \right),
\end{aligned}
\end{equation}
where $E_l = h v / l \approx 300 \, \theta^\circ \left( v / v_G \right) \,\,$meV, with $v_G = 10^6$~m~s$^{-1}$ the Fermi velocity of graphene, is the energy scale of the network. The network modes pick up a dynamical phase as they propagate along a link with velocity $v$, which we assume is equal for the two valley Hall states. For a clean network with translational invariance, the Bloch theorem gives \cite{Efimkin2018,Pal2019} 
\begin{equation} \label{eq:bloch}
\begin{pmatrix} b_{1m-1n-1} \\ b_{2m+1n} \\ b_{3mn+1} \end{pmatrix}_{\bm k} = \mathcal M(\bm k) \begin{pmatrix} b_{1mn} \\ b_{2mn} \\ b_{3mn} \end{pmatrix}_{\bm k},
\end{equation}
with $\mathcal M(\bm k) = \textrm{diag} \left( e^{ik_3}, e^{ik_1}, e^{ik_2} \right)$ where $k_j = \bm k \cdot \bm l_j$ ($j=1,2,3$) and $\bm l_3 = -(\bm l_1+\bm l_2)$. If we now combine Eqs.\ \eqref{eq:Smat} and \eqref{eq:bloch}, we obtain
\begin{equation} \label{eq:bands}
\left[ \mathcal M(\bm k) \otimes \mathds 1_2 \right] \mathcal S \, a_{\bm k} = e^{-i2\pi E/E_l} a_{\bm k},
\end{equation}
such that the network energy bands are given by the phase of the eigenvalues of $\left[ \mathcal M(\bm k) \otimes \mathds 1_2 \right] \mathcal S$.

In mTBG, the scattering parameters likely depend on the Fermi energy, as well as the interlayer bias, twist angle, etc., which alter the microscopic details of the chiral modes and the scattering nodes. In this work, we assume the $S$ matrix is independent of energy, which we estimate should hold for energy windows that are small compared to the local gap induced by the interlayer bias.

\subsection{Chiral zigzag modes and pseudo-Landau levels}

As a first approximation, we consider the case without forward scattering, i.e., $s_f=0$ in Eq.~\eqref{eq:C3}. This is a good approximation if the localization length of the chiral modes is not too small relative to the moir\'e scale. In this case, there can be a large wavefunction overlap between incoming and outgoing modes for deflections \cite{Qiao2014a}. Hence, we only need to determine $s_r$ subject to
$s_r s_r^* = 0$ and $s_r^\dag s_r + s_r^*( s_r )^t = \mathds 1_2$. These conditions are satisfied up to an overall phase by
\begin{equation}
s_r = \begin{pmatrix} e^{i\phi}\sqrt{P_{d1}} & \sqrt{P_{d21}} \\ -\sqrt{P_{d22}} & -e^{-i\phi}\sqrt{P_{d1}} \end{pmatrix},
\end{equation}
with $s_l=(s_r)^t$ and where $P_{d1}$ ($P_{d21}$ and $P_{d22}$) is the intrachannel (interchannel) deflection probability, where $2P_{d1}+P_{d21}+P_{d22}=1$ and $P_{d1}=\sqrt{P_{d21}P_{d22}}$. Hence, there are two parameters, e.g., $P_{d1}$ and $\phi$ with $0 \leq P_{d1} \leq 1/4$.
Equation \eqref{eq:bands} gives the secular equation
\begin{equation} \label{eq:secular}
\begin{aligned}
& 1 - \lambda^6 + \lambda^2 \left[ \lambda^2 \alpha(\bm k) - \alpha(\bm k)^* \right] \left( 1 - 4 P_{d1} \sin^2 \phi \right) \\
& \qquad + 2i \lambda^3 \left( 2 \sqrt{P_{d1}} \, \sin \phi \right)^3 = 0,
\end{aligned}
\end{equation}
with $\lambda=e^{i2\pi E/E_l} $ and $\alpha(\bm k) = e^{ik_1} + e^{ik_2} + e^{ik_3}$ which is invariant under $C_3$. Solving for $\lambda$ yields the network energy bands $E(\bm k)$. Since the energy enters via the dynamical phase, the spectrum is periodic and thus unbounded. In mTBG, the network exists only in a finite energy range where the gap is opened by the interlayer bias. We can simplify our analysis by defining $\sin \phi' = 2 \sqrt{P_{d1}} \sin \phi$ which has a real solution for $\phi'$ because $0 \leq P_{d1} \leq 1/4$. Hence, we find that Eq.~\eqref{eq:secular} is equivalent to $P_{d1}=P_{d21}=P_{d22}=1/4$ with $\phi \rightarrow \phi'$,
and we only have to consider this special case. In the following, we drop the prime for convenience. Furthermore, by inspection, we observe that all distinct cases are obtained for $\phi \in [0,\pi/2]$ since the substitution $\phi \rightarrow \pi - \phi$ leaves the secular equation invariant and $\phi \rightarrow -\phi$ is equivalent to $\lambda \rightarrow - \lambda$ which is a constant energy shift. 

In two limiting cases, we find that Eq.~\eqref{eq:secular} yields an analytical solution: $\phi=0$ and $\phi=\pi/2$. For $\phi=0$, the network bands are given by
\begin{equation} \label{eq:spectrum1}
E_{nj}(\bm k) = \frac{\hbar v}{2l} \left( 2 \pi n + k_j \right),
\end{equation}
where $n$ is an integer, $j=1,2,3$, and which is shown in Fig.~\ref{fig:bands1}(a). The network modes are given in this case by three chiral modes with velocity $v/2$ that propagate in the $\bm l_j$ directions. When $\phi > 0$, these modes hybridize [Fig.~\ref{fig:bands1}(b)]. However, certain crossings at the $\bar \Gamma$, $\bar K$, and $\bar K'$ points in the moir\'e Brillouin zone (MBZ) are symmetry-protected by $C_2T$ and $C_3$ symmetry. At the $\bar \Gamma$ point, the secular equation Eq.~\eqref{eq:secular} reduces to
\begin{equation} \label{eq:secular2} 
\left( \lambda^2 + 2 i \lambda \sin \phi - 1 \right) \left( \lambda^2 - i \lambda \sin \phi - 1 \right)^2 = 0, \\
\end{equation}
such that there always is a doubly-degenerate mode at the $\bar \Gamma$ point which becomes triply degenerate for the special case $\phi=0$. 
Increasing $\phi$ further, an indirect band gap opens at $\phi=\pi/6$, see Fig.~\ref{fig:bands1}(c). Hence, in the absence of forward scattering, the network is metallic for $0 \leq \phi \leq \pi/6$ and gapped for $\pi/6 \leq \phi \leq \pi/2$. 
Furthermore, for $\phi=\pi/2-\epsilon$, we find in lowest order of $\epsilon$
\begin{equation}
E_{nj\pm} \simeq \frac{\hbar v}{l} \left[ 2\pi n + \frac{\pi}{6} (4j-7) \pm \frac{\epsilon}{3} \left| f (\bm k-\bm k_{0j} ) \right| \right], 
\end{equation} 
where $\bm k_{0j} =(4\pi/3l) (j-1) \bm e_x$ and $f(\bm k) = 1 + e^{i\bm k \cdot \bm l_1} + e^{-i\bm k \cdot \bm l_2}$. At $\phi=\pi/2$ ($\epsilon=0$) there are three doubly-degenerate flatbands per period $E_l$ that are equally separated by $E_l/3$. For finite $\epsilon$, the degeneracy is lifted and each pair of network bands resembles those of graphene with only nearest-neighbor hopping, see Fig.~\ref{fig:bands1}(d). However, in this case the spectrum of each pair is shifted in momentum space by $\bm k_{0j}$.
\begin{figure}
\centering
\includegraphics[width=\linewidth]{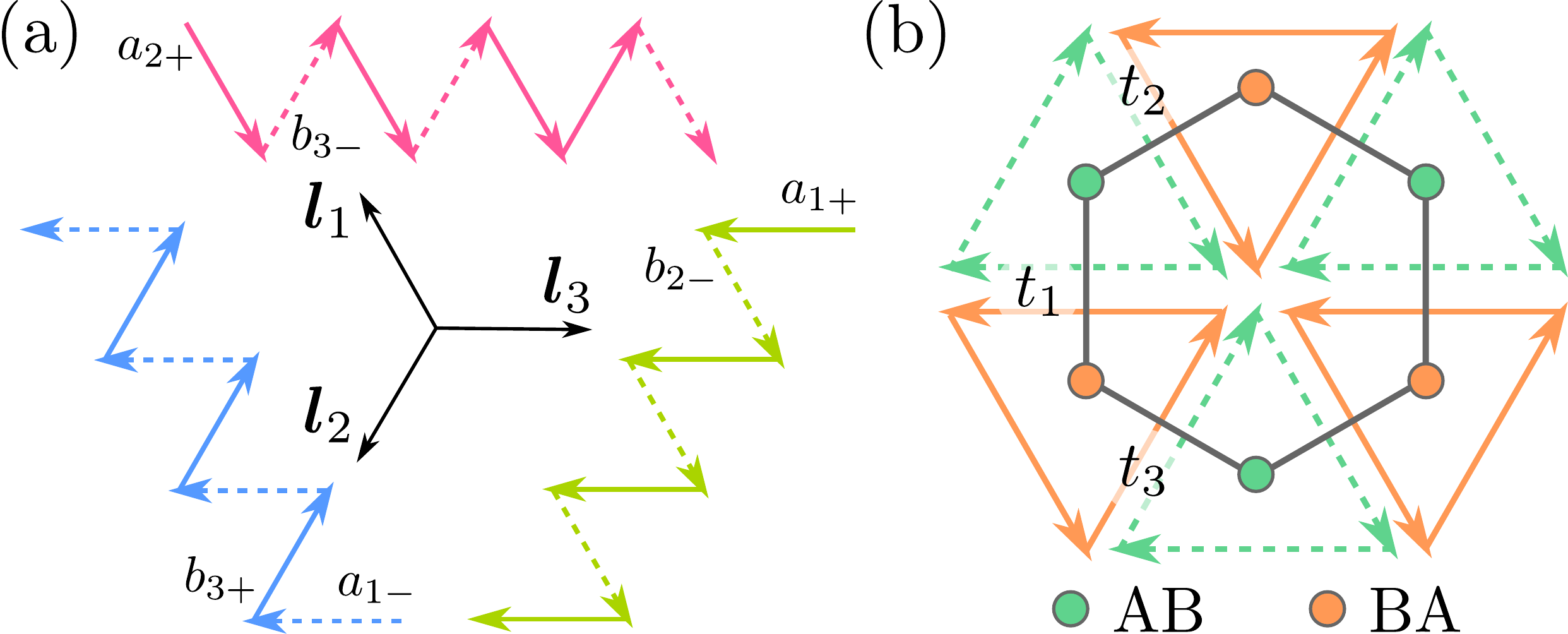}
\caption{(a) Triplet of 1D chiral zigzag modes ($\phi = 0$) along directions $\bm l_j$ ($j=1,2,3$) where solid (dashed) lines correspond to $a_+$ ($a_-$) superpositions of valley Hall states along the same link [Eq.\ \eqref{eq:sup}]. (b) Pseudo-Landau levels ($\phi = \pi/2$) with the effective honeycomb lattice superimposed.}
\label{fig:noforward}
\end{figure}

To understand these results physically, we perform a basis transformation, $U = \mathds 1_3 \otimes e^{-i\pi \sigma_y/4} e^{i \phi \sigma_z/2}$. The new basis corresponds to superpositions of the two valley Hall states on the same link,
\begin{equation} \label{eq:sup}
a_\pm = \frac{1}{\sqrt{2}} \left( a e^{i \phi / 2} \mp a' e^{-i \phi / 2} \right),
\end{equation}
and we obtain
\begin{equation} \label{eq:Snew}
U \mathcal S U^{-1} = \begin{pmatrix} \tilde s_f & \tilde s_l & \tilde s_r \\ \tilde s_r & \tilde s_f & \tilde s_l \\ \tilde s_l & \tilde s_r & \tilde s_f \end{pmatrix},
\end{equation}
where $\tilde s_f$ remains zero, and
\begin{equation}
\tilde s_r = \begin{pmatrix} 0 & \cos \phi \\ 0 & i \sin \phi \end{pmatrix}, \qquad
\tilde s_l = \begin{pmatrix} i \sin \phi & 0 \\ \cos \phi & 0 \end{pmatrix}.
\end{equation}

Thus, in the new basis, we see that for $\phi=0$ there are only interchannel deflections to the right (left) for $a_+$ ($a_-$) modes. This gives rise to a chiral zigzag motion in the network, as shown in Fig.~\ref{fig:noforward}(a), where $a_+$ ($a_-$) modes correspond to solid (dashed) arrows. We call these modes chiral zigzag (ZZ) modes \cite{Tsim2020,DeBeule2020a}. Moreover, because of the ZZ motion, they effectively have a velocity of $v/2$ as it takes twice as long compared to direct propagation between two nodes. In the flatband limit ($\phi=\pi/2$) we find that the network modes in the new basis form two sets of decoupled trimers, given by clockwise (counterclockwise) triangular orbits around AB (BA) regions for $a_-$ ($a_+$) modes. Hence, the network is localized and we obtain flatbands which are known in the literature as pseudo-Landau levels \cite{Ramires2018}. Note that $C_2T$ symmetry is conserved on the whole, because the orbits rotate in opposite directions. For a given orientation, we obtain three pseudo-Landau levels per network period as each orbit consists of three parts, giving three possible superpositions with a different energy. Furthermore, pseudo-Landau levels of opposite orientation can be thought of as two sublattices of an effective honeycomb lattice, see Fig.~\ref{fig:noforward}(b). In this way, the spectrum for $\phi=\pi/2-\epsilon$ is understood in terms of nearest-neighbor hoppings $t_1=\epsilon/3$ and $t_2=t_3^*=e^{i2\pi/3} \epsilon/3$ on a honeycomb lattice, where $\epsilon$ is the amplitude for a chiral ZZ process.
\begin{figure}
\centering
\includegraphics[width=\linewidth]{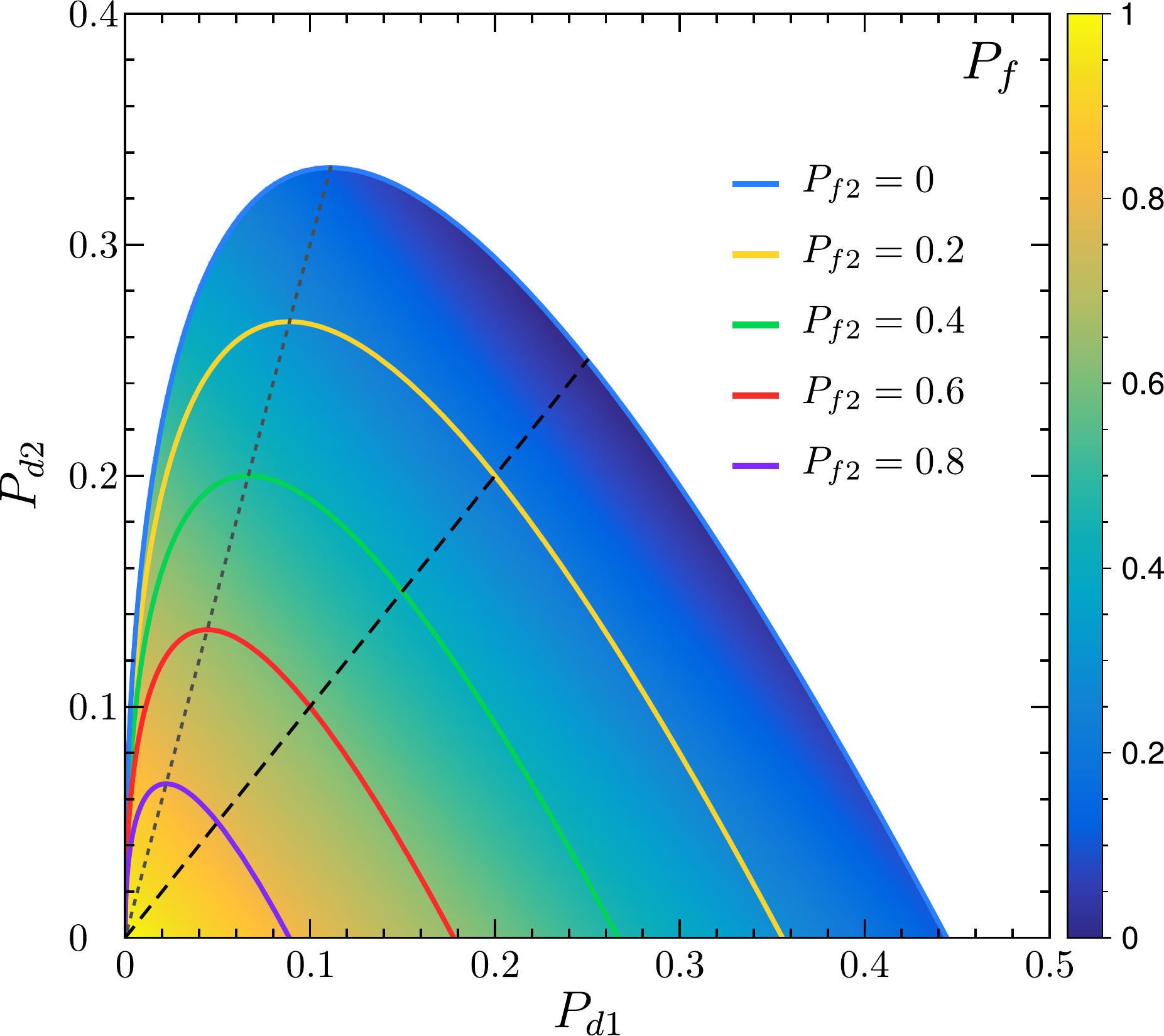}
\caption{Regions in the $(P_{d1},P_{d2})$ plane consistent with unitarity as well as $C_3$ and $C_2T$ symmetry. For a given value of $P_{f2}$, the allowed region corresponds to the area bounded by the curve and the $P_{d1}$ axis. The density plot shows the total forward scattering probability $P_f=P_{f1}+P_{f2}$ and the dashed (dotted) line corresponds to $P_{d2}=P_{d1}$ ($P_{d2}=3P_{d1}$) where the latter gives the maximal value of $P_{d2}$ for a given $P_{f2}$.}
\label{fig:phase2channel}
\end{figure}

In the absence of forward scattering, we conclude that the network model already captures two phenomena which were known from band-structure calculations: chiral zigzag modes \cite{Fleischmann2020,Tsim2020} and pseudo-Landau levels \cite{Ramires2018}. Next, we demonstrate how the network model can be used to understand interference oscillations in magnetotransport in terms of couplings between zigzag modes.

\subsection{Coupling of zigzag modes}

\begin{figure}
\centering
\includegraphics[width=\linewidth]{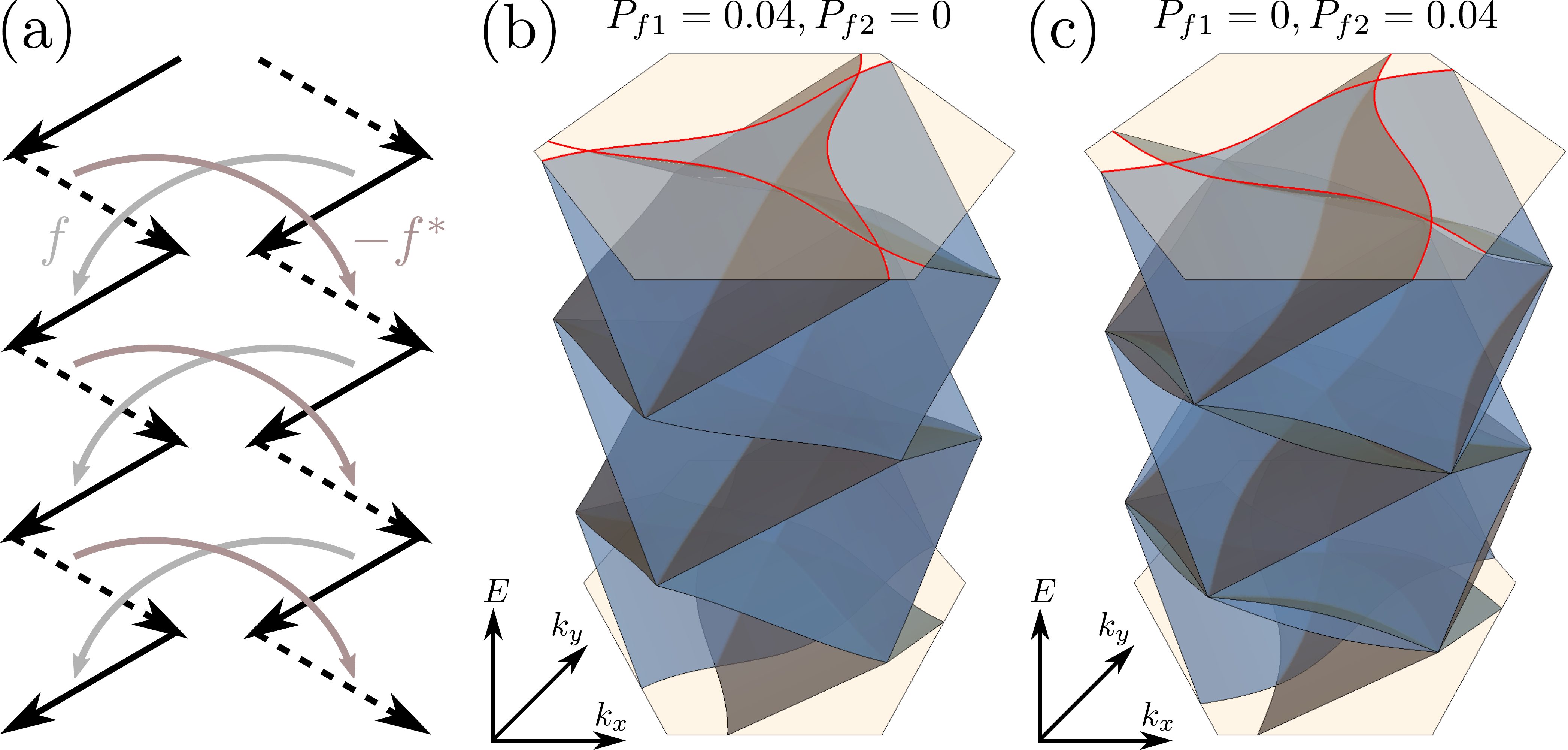}
\caption{(a) Schematic showing scattering processes between parallel ZZ channels. (b)--(c) Network energy bands in the MBZ for $\Delta=\phi=0$ with (b) $P_{f1} = 0.04$ and $P_{f2} = 0$, and (c) $P_{f1} = 0$ and $P_{f2} = 0.04$.}
\label{fig:bands2}
\end{figure}
In the limit of tiny twist angles $\theta \sim 0.1^\circ$, the localization length of the chiral modes decreases as the domain walls become more sharply defined due to lattice relaxation \cite{Walet2019,Yoo2019}. Hence, we expect that forward scattering becomes more likely. Therefore, we consider the following parameterization \cite{DeBeule2020a}
\begin{align}
s_f & = \begin{pmatrix} e^{i(\phi+\chi)}\sqrt{P_{f1}} & -\sqrt{P_{f2}} \\ -\sqrt{P_{f2}} & -e^{-i(\phi+\chi)}\sqrt{P_{f1}} \end{pmatrix}, \label{eq:sf} \\
s_r & = \begin{pmatrix} e^{i\phi}\sqrt{P_{d1}} & \sqrt{P_{d2}} \\ -\sqrt{P_{d2}} & -e^{-i\phi}\sqrt{P_{d1}} \end{pmatrix}, \label{eq:sr} 
\end{align}
with $s_l=(s_r)^t$ and where $P_{f1}$ ($P_{f2}$) is the probability for intrachannel (interchannel) forward scattering, and $P_{d1}$ ($P_{d2}$) is the probability for intrachannel (interchannel)  deflections. Here, the phase shift $\phi$ is an independent real parameter. The $S$ matrix is unitary for $2(P_{d1}+P_{d2})+P_{f1}+P_{f2}=1$ and $\cos \chi = \left( P_{d2} - P_{d1} \right)/2\sqrt{P_{f1} P_{d1}}$, where we take $\chi \geq 0$. Moreover, $\chi$ has to be real which implies $2\sqrt{P_{f1}P_{d1}} \geq \left| P_{d2} - P_{d1} \right|$. Hence, we obtain an upper bound on $P_{d2}$ as illustrated in Fig.~\ref{fig:phase2channel}, where we show the allowed regions in the $(P_{d1},P_{d2})$ plane for different values of $P_{f2}$ and where we have superimposed the value of $P_f=P_{f1}+P_{f2}$. These regions are bounded by
\begin{equation}
P_{d2} \leq 2 \sqrt{P_{d1} \left( 1-P_{f2} \right)} - 3P_{d1},
\end{equation}
which is independent of $\phi$. 
Note that this parameterization is not the most general. An explicit construction of the general $S$ matrix is given in App.\ \ref{app:model}. However, we believe this parameterization suffices to capture the network physics in mTBG. Henceforth, we consider four independent scattering parameters that can be chosen as $\phi$, $P_{f1}$, $P_{f2}$, and $\Delta=P_{d1}-P_{d2}$. 

To investigate the effect of forward scattering on the chiral zigzag modes, we make the same basis transformation as in Eq.~\eqref{eq:Snew}. In the new basis, we find that the $S$ matrix can be written as
\begin{equation}
\tilde s_f = \begin{pmatrix} f & -ig \\ ig^* & -f^* \end{pmatrix}, \,\,\,\,\,
\tilde s_{r,l} = \begin{pmatrix} i \delta_\mp \sin \phi & \delta_\pm \cos \phi \\ \delta_\mp \cos \phi & i \delta_\pm \sin \phi \end{pmatrix},
\end{equation}
with
\begin{align}
f & = \sqrt{P_{f2}} \, \cos \phi + i \sqrt{P_{f1}} \, \sin ( \phi + \chi), \\
g &= \sqrt{P_{f2}} \, \sin \phi + i \sqrt{P_{f1}} \, \cos \left( \phi + \chi \right), \\
\delta_\pm & = \sqrt{P_{d1}} \pm \sqrt{P_{d2}}.
\end{align}
In the ZZ regime ($\phi=\delta_-=0$) we have $\chi=\pi/2$ and $g=0$, such that only parallel zigzag channels are coupled due to intrachannel forward scattering in the $a_\pm$ basis with probability $|f|^2$, which is illustrated in Fig.~\ref{fig:bands2}(a). In this case, the spectrum becomes $(j=1,2,3)$
\begin{equation} \label{eq:bands2}
\begin{aligned}
& E_{nj\pm}(\bm k) = \\
& \quad \frac{\hbar v}{l} \left[ 2 \pi \left( n + \frac{1 \pm 1}{4} \right) + \frac{k_j}{2} \pm \arcsin F_j(\bm k) \right],
\end{aligned}
\end{equation}
where
\begin{equation}
F_j = \sqrt{P_{f1}} \, \cos q_j - \sqrt{P_{f2}} \, \sin q_j,
\end{equation}
where $q_j = k_j / 2 + k_{j+1}$ with $k_4=k_1$ and which is shown in Figs.\ \ref{fig:bands2}(b) and (c). Note that the accidental triple degeneracy at the $\bar \Gamma$, $\bar K$, and $\bar K'$ points is not lifted by forward scattering, such that it cannot open a gap. Indeed, forward scattering only results in an overall energy shift at the high-symmetry points, as $F_j(\bar \Gamma) = \sqrt{P_{f1}}$ and $F_1(\pm \bar K) = F_2(\pm \bar K) = -F_3(\pm \bar K) = -\sqrt{P_{f1}}$. Furthermore, we observe that the Fermi surface becomes warped depending on the type of forward scattering, see Fig.~\ref{fig:bands2}. For example, in the absence of interchannel forward scattering ($P_{f2}=0$) the bands are symmetric in $k_y$ which follows from $F_1(k_x,-k_y)=F_2(k_x,k_y)$ and $F_3(k_x,-k_y)=F_3(k_x,k_y)$. 

\begin{figure}
\centering
\includegraphics[width=\linewidth]{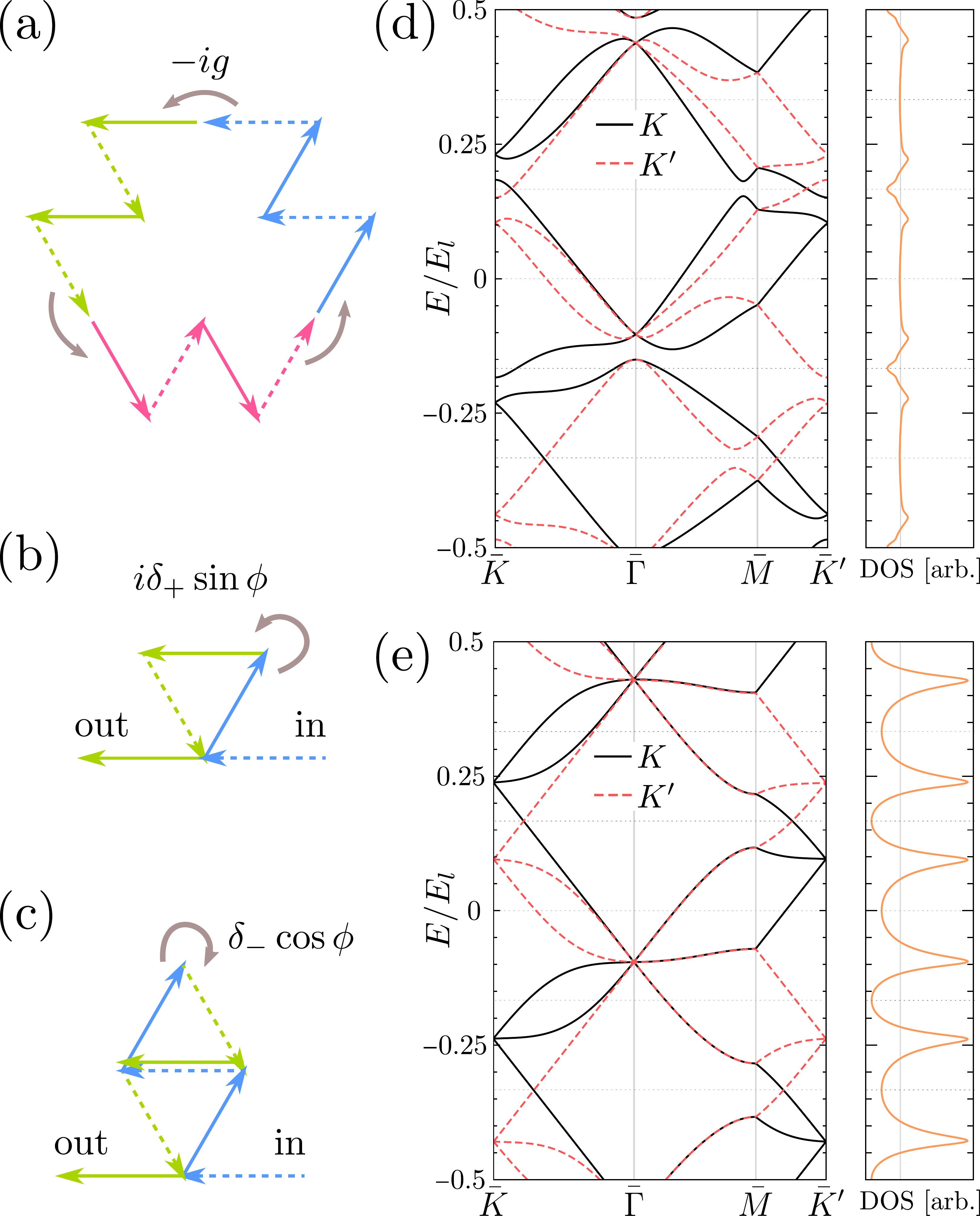} 
\caption{(a)--(c) Scattering between different chiral zigzag branches. (a) Interchannel forward scattering where $a_+$ ($a_-$) modes are solid (dashed) arrows. (b) Intrachannel deflections. (c) Interchannel deflections that are opposite to zigzag processes. (d) Network bands and density of states (DOS) for $\phi = 0.2$, $\Delta = P_{d1} - P_{d2} = 0$, and $P_{f1} = P_{f2} = 0.05$ and (e) for $\phi = 0$, $\Delta = 0.4$, $P_{f1} = 0.1$, and $P_{f2} = 0$. The gray line in the DOS corresponds to the constant DOS $\left( 4 / \mathcal A \right) \left( 6 / E_l \right)$ in the chiral ZZ regime, where $\mathcal A$ is the moir\'e cell area.}
\label{fig:bands3}
\end{figure}

In general, however, the three chiral zigzag channels are coupled through several processes, which are illustrated in Fig.~\ref{fig:bands3}(a), (b), and (c). For example, one such process is due to interchannel forward scattering in the $a_\pm$ basis with probability $|g|^2$ [Fig.~\ref{fig:bands3}(a)]. 
Consequently, the bands belonging to different zigzag branches hybridize except at the $\bar \Gamma$, $\bar K$, and $\bar K '$ points of the MBZ where the crossing is protected by $C_3$ and $C_2T$ symmetry, see Fig.~\ref{fig:bands3}(d). Moreover, for $\phi=0$ there is always an accidental triply degenerate crossing at high-symmetry points regardless of the value of $\Delta$, which prevents a gap from being opened, as shown in Fig.~\ref{fig:bands3}(e).

The nesting between the Fermi surfaces of opposite valleys, which was first investigated with the continuum model \cite{Fleischmann2020}, can thus be understood from the 1D nature of the zigzag channels. 
By comparing the network energy bands to microscopic band-structure calculations \cite{Fleischmann2020,Tsim2020}, we estimate that $\phi \approx \Delta \approx0$ for mTBG and that forward scattering increases in the presence of lattice relaxation, as evidenced by the deformation of the Fermi contour.

\section{Transport in the presence of external fields} \label{sec:transport}

We now study electronic transport through the network in the presence of a magnetic field $\bm B = B \bm e_z$ perpendicular to the graphene layers, as well as a uniform in-plane electric field $\bm{\mathcal E} = \mathcal E \bm e_x$. 

\begin{figure}
\centering
\includegraphics[width=\linewidth]{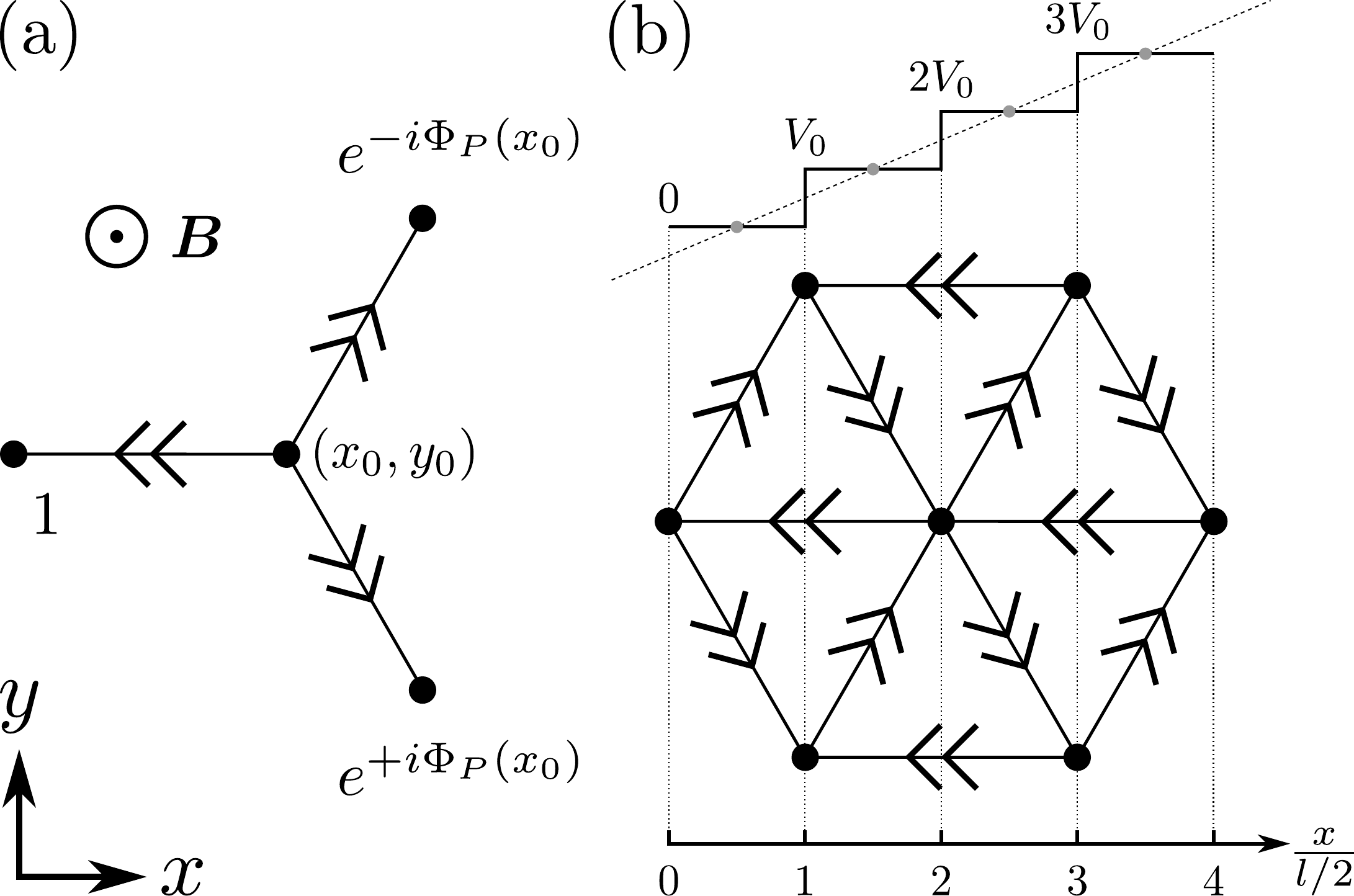}
\label{fig:fields}
\caption{(a) Peierls phase in the Landau gauge $\bm A = Bx \bm e_y$ accumulated after propagation along links. (b) Staircase potential with step height $V_0$ along the $x$ direction.}
\end{figure}

In the gauge $\bm A = Bx \bm e_y$, a Peierls phase $\pm \Phi_P (x)$ is accumulated along a downward/upward diagonal link starting at a node with horizontal position $x$, with
\begin{equation}
\Phi_P (x) = \frac{\pi \Phi}{\Phi_0} \left( \frac{x}{l/2} + \frac{1}{2} \right),
\end{equation}
where $\Phi=B \mathcal A$ is the flux through a moir\'e cell, $\mathcal A = \sqrt{3} \, l^2/2$ is the moir\'e cell area, and $\Phi_0=h/e$. In this gauge, no Peierls phase is accumulated along horizontal links. This is illustrated in Fig.~\ref{fig:fields}(a). 
We introduce the in-plane electric field in the network model via the dynamical phase \cite{Cedzich2013} with the substitution $E \rightarrow E - V(x)$ where $V(x) = e\mathcal E \left( x -  l / 4 \right)$. Note that the total phase after propagation, e.g., along a diagonal link, 
\begin{equation}
\frac{1}{\hbar v} \int ds \, V = \frac{2}{\hbar v} \int_{x_0}^{x_0+l/2} dx \, V(x) = \frac{2\pi x_0}{l/2} \frac{V_0}{E_l},
\end{equation}
with $V_0 = e\mathcal E l/2$, is the same for the staircase potential
\begin{equation}
V(x) = V_0 \sum_n n \left[ \theta(x-nl/2) - \theta(x-(n+1)l/2) \right],
\end{equation}
as shown in Fig. \ref{fig:fields}(b). 

We want to emphasize that both the magnetic and in-plane electric field break the symmetries of the triangular network, i.e., the former breaks time-reversal symmetry and the latter breaks rotation symmetry. Hence, there are less constraints on the $S$ matrix in general. We therefore consider the regime where the magnetic length is large compared to the dimensions of the scattering region. We further assume that the variation of the potential in the scattering region is small on the scale of the confinement energy. In this regime, we expect that the symmetries are conserved locally, such that the $S$ matrix approximately maintains its form dictated by $C_3$ and $C_2T$ symmetry. At minimal twists $\theta \sim 0.1^\circ$, the diameter of the scattering region is a fraction $\alpha(\theta) \ll 1$ of the moir\'e length $l$ due to lattice relaxation \cite{Nam2017,Walet2019,Yoo2019}. Hence, our assumption is reasonable for $l_B > \alpha l$, where $l_B$ is the magnetic length, which is equivalent to $\Phi/\Phi_0 < \alpha^{-2}$, as well as $V_0/E_l < \alpha^{-1}$. Note that it is still possible that the $S$ matrix gains an $x$ dependence due to $V(x)$. We do not consider such a dependence in this work.

\subsection{Four-terminal setup}

\begin{figure}
\centering
\includegraphics[width=\linewidth]{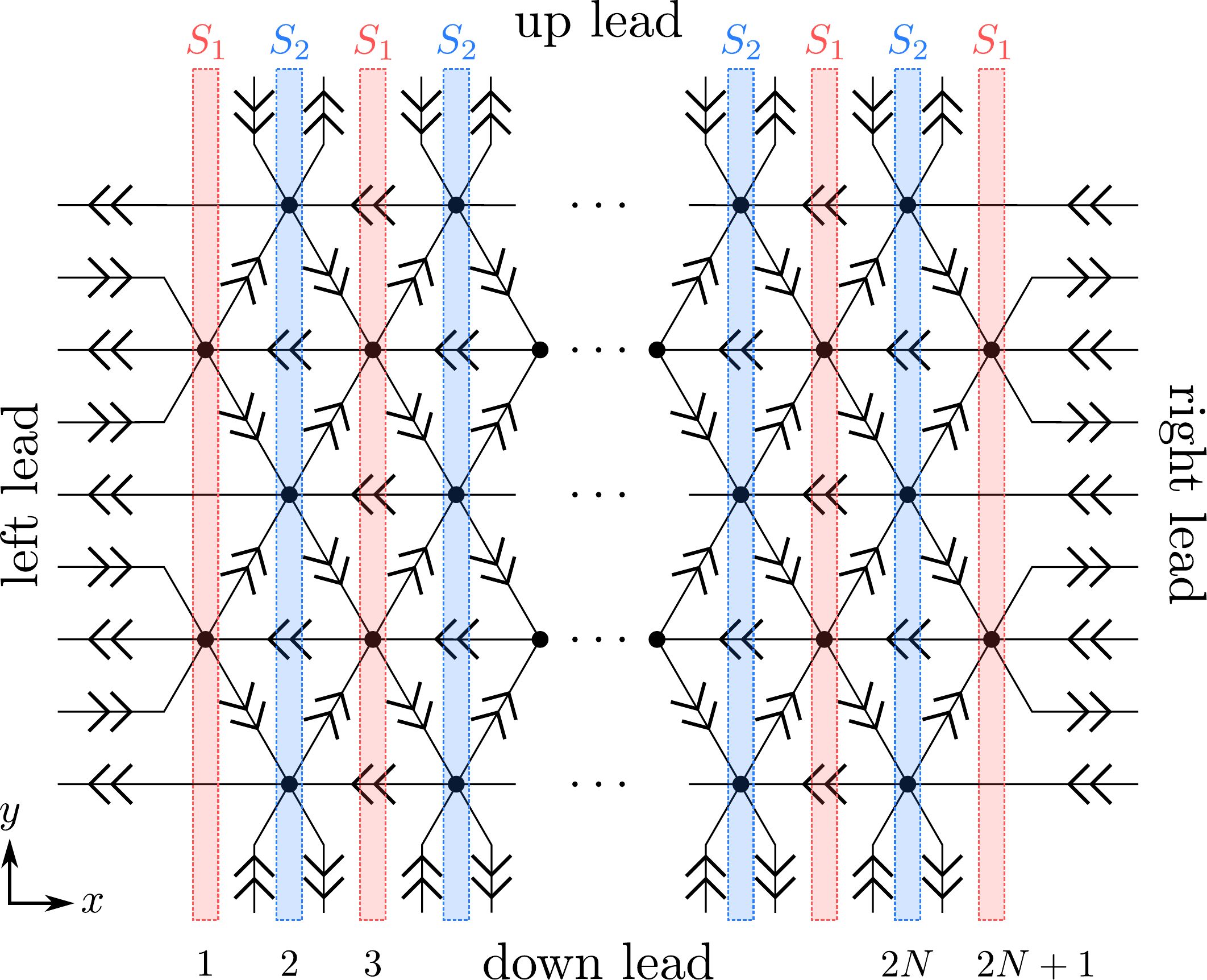}
\label{fig:4terminal}
\caption{Four-terminal setup with width $W=\sqrt{3}Ml$ and length $L=Nl$, shown here for $M=2$.}
\end{figure}

We calculate the transmission functions in a four-terminal setup with length $L=Nl$ and width $M=\sqrt{3} M l$, which is shown in Fig.~\ref{fig:4terminal}. Hence, we need to calculate the total $S$ matrix of the open system. To this end, we divide the network along the $x$ direction in $2N+1$ transverse slices, labeled by $n=1,\ldots,2N+1$. If each node is equivalent, odd (even) numbered slices always have $S$ matrix $S_1$ ($S_2$), as illustrated in Fig.~\ref{fig:4terminal}. These $S$ matrices are explicitly given in App.\ \ref{app:4terminal}. Hence, the total $S$ matrix can be written as
\begin{equation}
S = \underbrace{S_1 \times S_2 \times \cdots \times S_1 \times S_2 \times S_1}_{2N+1~\textrm{factors}},
\end{equation}
where the operation $\times$ refers to combining $S$ matrices and which is explicitly defined in App.\ \ref{app:4terminal}. The total $S$ matrix then becomes
\begin{align}
& \quad {\small \begin{matrix} $\,4M$ & $\,\,4M+2$ & $\,\,2N$ & $\,\,\,2N$ \end{matrix}} \nonumber \\
S = & \begin{pmatrix}
r_L & \hphantom{d}t_{LR}\hphantom{d} & t_{LU} & t_{LD} \\
t_{RL} & r_R & t_{RU} & t_{RD} \\
t_{UL} & t_{UR} & r_U & t_{UD} \\
t_{DL} & t_{DR} & t_{DU} & r_D
\end{pmatrix} {\small \begin{matrix} $4M+2$ \\[.5mm] $4M$ \\[.5mm] $2N$ \\[.5mm] $2N$ \end{matrix}}, \label{eq:totalS}
\end{align}
where $t_{\alpha\beta}$ are matrices that contain scattering amplitudes from lead $\beta$ to lead $\alpha$ with $r_\alpha = t_{\alpha\alpha}$, and where the labels $L$, $R$, $U$, and $D$ correspond to the left, right, up, and down leads, see Fig.~\ref{fig:4terminal}. Here, we show the number of rows and columns above and on the right side of the matrix, respectively. Note that Eq.~\eqref{eq:totalS} gives the $S$ matrix for one valley, and that $S_K(\Phi) = \left[ S_{K'}(-\Phi) \right]^t$. The transmission functions then become
\begin{equation}
T_{\alpha\beta}(\Phi) = \sum_{\tau=K,K'} \textrm{Tr} \big( t_{\alpha\beta}^\dag t_{\alpha\beta}  \big)_\tau = T_{\beta\alpha}(-\Phi),
\end{equation}
where the transmission of different valleys is generally not equal in the four-terminal setup. Remember that the total network (disregarding spin) consists of two decoupled networks, one for each valley, related by time reversal.

In the Landauer-B\"uttiker formalism, the current at lead $\alpha$ is given in linear response by \cite{Buttiker1986,Buttiker1988}
\begin{equation} \label{eq:current}
I_\alpha = \sum_{\beta} G_{\alpha\beta} \left( V_\alpha - V_\beta \right),
\end{equation}
where $V_\alpha$ is the voltage at lead $\alpha$ and
\begin{equation} \label{eq:GT} 
G_{\alpha\beta} = \frac{2e^2}{h} \int dE \, T_{\alpha\beta} \left( - \frac{\partial f_0}{\partial E} \right) \stackrel{T \rightarrow 0}{\rightarrow} \frac{2e^2}{h} T_{\alpha\beta}(E_F),  
\end{equation}
where $f_0$ is the Fermi-Dirac distribution, $E_F$ the Fermi level, and $\alpha, \beta = L, R, U, D$. We then determine the longitudinal conductances $G_{xx}$ and $G_{yy}$, as well as the transverse responses $G_{xy}$ and $G_{yx}$ from Eq.~\ref{eq:current} for the case where a current is applied between opposite leads ($I_R = -I_L$ and $I_U = -I_D$). More details can be found in App.\ \ref{app:4terminal}.

Finally, we briefly address the symmetry constraints on the conductivity tensor $\sigma$. The Onsager reciprocal relations are given by $\sigma_{ij}(\Phi) = \sigma_{ji}(-\Phi)$ ($i,j=x,y$) or $\sigma(\Phi) = \sigma(-\Phi)^t$ such that $\sigma$ is symmetric when time-reversal symmetry is preserved. In addition, for the bulk network $C_3$ symmetry implies $\sigma = R(2\pi/3) \sigma R(-2\pi/3)$ such that $\sigma_{xx}=\sigma_{yy}$ and $\sigma_{xy}=-\sigma_{yx}$ in the absence of an in-plane electric field. The latter also hold within a single valley as $C_3$ conserves the valley. In general, we can write $\sigma_{xy/yx} = \sigma_S \pm \sigma_H$ where $\sigma_H = \left( \sigma_{xy} - \sigma_{yx} \right) / 2$ is the Hall response and $\sigma_S = \left( \sigma_{xy} + \sigma_{yx} \right) / 2$. While the Hall response $\sigma_H$ vanishes in the presence of time-reversal symmetry, the symmetric (Drude) part $\sigma_S$ can be finite even when time-reversal symmetry is preserved in the absence of rotation symmetry \cite{Ortix2021,Wu2017}. Since the four-terminal setup breaks $C_3$ symmetry, we find that $\sigma_S$ is generally nonzero in the coordinate system of Fig.~\ref{fig:4terminal}. Furthermore, for a given value of $\sigma_S$ one can always find new coordinates $(x',y')^t = O (x,y)^t$ with $O O^t = 1$ such that $\sigma_S \rightarrow 0$, while $\sigma_H \rightarrow \det O \, \sigma_H$ transforms as a pseudoscalar.

\subsection{Chiral zigzag regime}

\begin{figure}
\centering
\includegraphics[width=\linewidth]{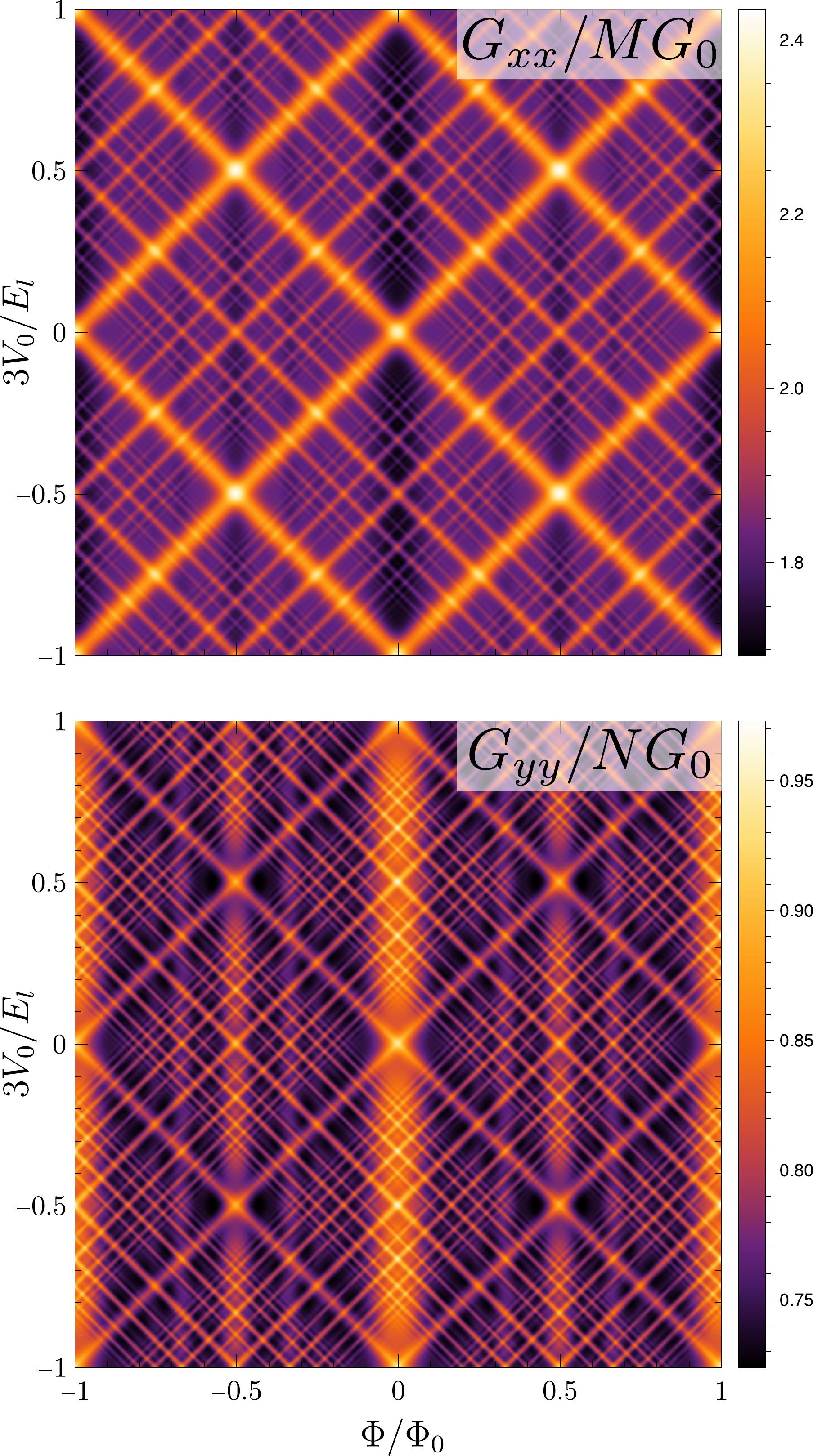}
\label{fig:condZZ}
\caption{Zero-temperature conductance $G_{xx}$ and $G_{yy}$ of the four-terminal setup with $N=M=10$ as a function of the magnetic flux $\Phi$ per moir\'e cell and the slope of the scalar potential $V(x) = 2V_0x/l$. Here, $G_0=4e^2/h$ and the scattering parameters are $\phi=\Delta=0$ and $P_{f1}=P_{f2}=0.4$.}
\end{figure}

In the chiral ZZ regime, different ZZ branches are decoupled ($\phi=\Delta=0$) and the network corresponds to three decoupled quasi-1D systems (for each valley and spin). We show the longitudinal conductance in this regime in Fig.\ \ref{fig:condZZ} as a function of the magnetic flux through a moir\'e cell $\Phi$ and the potential step height $V_0$.
We find Aharonov-Bohm (A-B) resonances that are split in the presence of an in-plane electric field. Since each ZZ branch propagates in a different direction, they accumulate a different phase in the potential $V(x)$, while the phase due to the magnetic field is the same for every branch. Hence, A-B resonances originating from different ZZ branches are affected differently by the in-plane electric field.
We demonstrate this explicitly for a small system for which the transmission can be calculated analytically. Because the ZZ branches are decoupled, the total transmission function can be written as $T = T_1 + T_2 + T_3$, where $T_j$ is the transmission of the ZZ branch that propagates in the $\bm l_j$ direction $(j=1,2,3)$. 
\begin{figure}
\centering
\includegraphics[width=\linewidth]{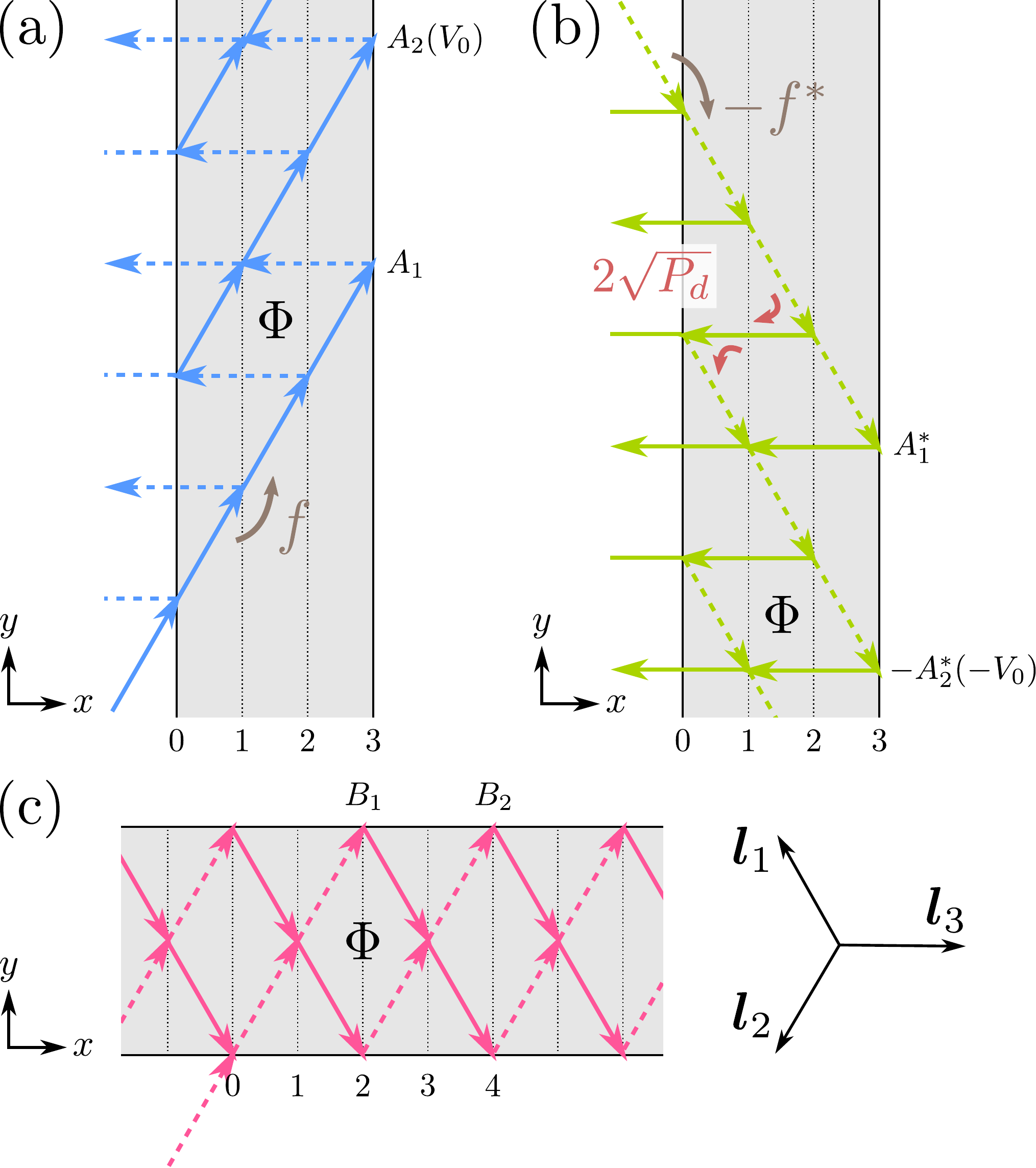}
\caption{Network strip of length $L=3l/2$ and infinite width, where we show the amplitudes of some of the shortest Feynman paths in the chiral zigzag regime ($\phi=\Delta=0$) for the $\bm l_1$ (a), $\bm l_2$ (b), and $\bm l_3$ (c) zigzag branches.}
\label{fig:an}
\end{figure}

Consider a network strip of length $L=3l/2$ and infinite width. The amplitudes of Feynman paths for the $\bm l_1$ branch, illustrated in Fig.~\ref{fig:an}(a), are given up to an overall phase by
\begin{align}
A_1 & = f^4, \\
A_2 & = f^4 (8P_df) \cos \left[ \pi \left( \frac{\Phi}{\Phi_0} + \frac{3V_0}{E_l} \right) \right], \\
& \,\,\, \vdots \nonumber \\
A_{n+1} & = f^4 (8P_d f)^n \cos^n \left[ \pi \left( \frac{\Phi}{\Phi_0} + \frac{3V_0}{E_l} \right) \right],
\end{align}
where the index labels the path length $3nl$, such that per transverse unit cell
\begin{align}
T_1(V_0) & = \sum_{n=1}^\infty |A_n|^2 \\
&  = \frac{P_f^4}{1 - 4 (1-P_f)^2 P_f \cos^2 \left[ \pi \left( \frac{\Phi}{\Phi_0} + \frac{3V_0}{E_l} \right) \right]},
\end{align}
and $T_2(V_0)=T_1(-V_0)$, see Fig.~\ref{fig:an}(b). Also, $T_3=2$ since there are two incoming modes per transverse unit cell that belong to the $\bm l_3$ branch, which are always transmitted as they are chiral in the transport direction. The total transmission becomes
\begin{equation}
T = 2 + T_1(V_0) + T_1(-V_0),
\end{equation}
which is independent of energy. This follows from the fact that interfering paths in the chiral ZZ regime always have the same length, and therefore the dynamical phase enters as an overall phase factor in the amplitudes. In turn, this implies that the A-B resonances are not smeared out at finite temperatures \cite{DeBeule2020a,Virtanen2011}, see Eq.~\eqref{eq:GT}.

We see that the potential enters in the transmission through $\Phi/\Phi_0 \rightarrow \Phi/\Phi_0 \pm 3V_0/E_l$, which also holds in larger systems for an arbitrary Feynman path. The resonance condition is thus given by
\begin{equation} \label{eq:resonance}
\frac{\Phi}{\Phi_0}  \pm \frac{3V_0}{E_l} = \frac{n}{m},
\end{equation}
with $n$ and $m \neq 0$ coprime integers and where the $+$ ($-$) sign corresponds to the $\bm l_1$ ($\bm l_2$) ZZ branch. Hence, an in-plane electric field separates the A-B resonances of different ZZ modes. This gives rise to the fractal pattern in the conductance shown in Fig.~\ref{fig:condZZ}. Moreover, along any resonant path in the ($\Phi,V_0)$ plane, there are additional resonances when the resonance conditions of different ZZ branches are satisfied simultaneously.
\begin{figure*}
\centering
\includegraphics[width=\linewidth]{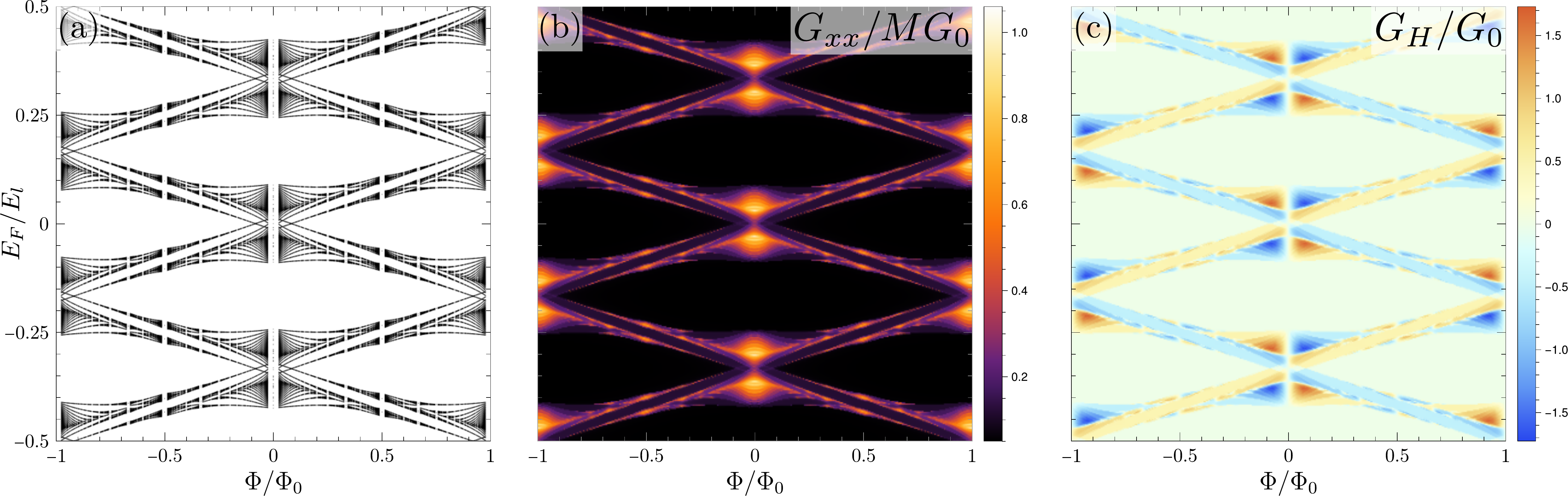}
\caption{(a) Hofstadter butterfly for $V_0 = 0$, $\phi=1$, and $P_{f1}=P_{f2}=\Delta=0$. (b) Longitudinal conductance and (c) Hall conductance for the four-terminal setup with $N=M=10$, for the same scattering parameters as in (a) with $G_0 = 4e^2/h$.}
\label{fig:cond2Da}
\end{figure*}

In the four-terminal setup, $G_{xx}$ and $G_{yy}$ also display broad antiresonances and resonances corresponding to the dark and bright vertical bands in Fig.~\ref{fig:condZZ}, respectively, that are independent of $V_0$. These are due to the $\bm l_3$ ZZ modes which propagate along the $x$ direction of the four-terminal setup shown in Fig.~\ref{fig:4terminal}. For this ZZ branch, the relative phase of interfering paths is unaffected by the potential $V(x)$. We illustrate this in Fig.~\ref{fig:an}(c), where we show a long network strip with width $W=\sqrt{3}l$. Here, the amplitudes of the three shortest paths for transmission in the $y$ direction of the $\bm l_3$ branch are
\begin{align}
B_1 & = (-f^*)^3, \\
B_2 & = (-f^*)^3 8P_d \cos \left( \pi \frac{\Phi}{\Phi_0} \right), \\
B_3 & = (-f^*)^3 4P_d \left[ 16P_d \cos^2 \left( \pi \frac{\Phi}{\Phi_0} \right) - 1 \right],
\end{align}
up to overall phases. Hence, the resonances are independent of $V_0$, giving rise to the background in $G_{yy}$. The antiresonances seen in $G_{xx}$ are understood in the same way, as some of the $\bm l_3$ ZZ modes which contribute a constant transmission in the $x$ direction, can now also transmit to the up and down leads.

From these results, we can make some predictions of magnetotransport in the ZZ regime at a potential hill that varies slowly on the moir\'e scale. Initially, the increase of the potential gives rise to resonances along lines in the $(\Phi,V_0)$ plane given by Eq.~\eqref{eq:resonance}. Resonant paths have either a positive or negative slope, depending on their respective ZZ branch. However, at the other end of the barrier, the sign of $V_0$ is changed. Therefore, resonances across the whole barrier appear only when two resonant paths of opposite slope intersect, which occurs only for rational numbers
\begin{equation}
\left( \frac{\Phi}{\Phi_0} , \frac{3V_0}{E_l} \right) = \left( \frac{n_1}{m_1} , \frac{n_2}{m_2} \right). 
\end{equation}

Finally, we find that the total transverse response $G_{xy} = \left( G_S + G_H \right) / 2$ vanishes in the ZZ regime. Within a single valley, the Hall response $G_H$ vanishes while the symmetric transverse response $G_S$ is nonzero in the presence of an in-plane electric field but opposite in each valley. This is allowed since both the four-terminal setup and the potential $V(x)$ break $C_3$ symmetry.

\subsection{Percolating regime}

When different zigzag branches become coupled ($\phi \neq 0$ or $\Delta \neq 0$) the network corresponds to a true two-dimensional percolating system. In this case, the presence of a perpendicular magnetic field gives rise to network Landau levels at small $\Phi/\Phi_0$ and Hofstadter physics when $\Phi / \Phi_0 \sim 1$, where
\begin{equation}
B \approx \frac{24 \Phi}{\Phi_0} \left( \theta^\circ \right)^2 \, \textrm{Tesla},
\end{equation}
such that the Hofstadter regime is experimentally accessible for minimal twist angles $\theta \sim 0.1^\circ$. For simplicity, we only consider the case without an in-plane electric field in this section $(V_0=0)$.

\begin{figure*}
\centering
\includegraphics[width=\linewidth]{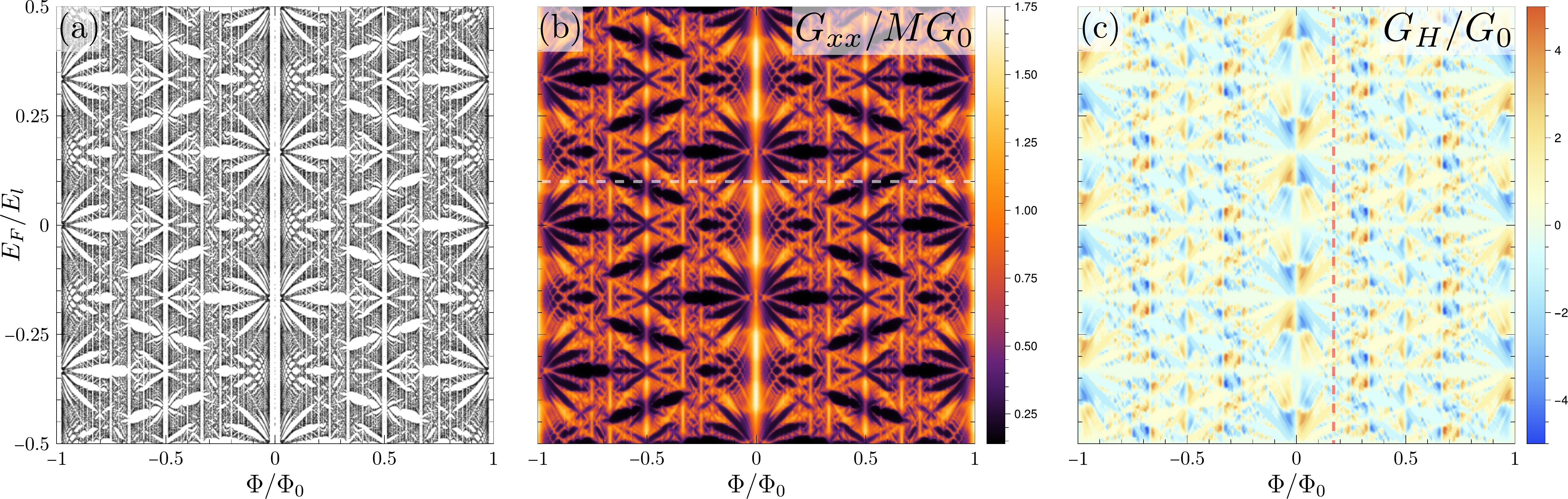}
\caption{(a) Hofstadter butterfly for $V_0=0$, $\phi=0.5$, $P_{f1}=P_{f2}=0.3$, and $\Delta=0$. (b) Longitudinal and (c) Hall conductance for the four-terminal setup with $N=M=20$, for the same scattering parameters as in (a) with $G_0 = 4e^2/h$.}
\label{fig:cond2Db}
\end{figure*}

In Figs.\ \ref{fig:cond2Da}(a) and \ref{fig:cond2Db}(a), we show the magnetic network bands at zero momentum, i.e., the Hofstadter butterfly (see App.\ \ref{app:hof}). The case shown in Fig.~\ref{fig:cond2Da} corresponds to a gapped network, which requires $\pi/6 \leq \phi \leq \pi/2$ without forward scattering \cite{DeBeule2020b}. Note that there are two bands of states with opposite spectral flow, indicative of a nontrivial topological phase \cite{Asboth2017,Chou2020}. Indeed, the two-channel triangular network in the presence of $C_2T$ symmetry hosts a valley anomalous Floquet insulator \cite{DeBeule2020b}. When the network is metallic at zero flux, several magnetic gaps develop at finite flux, as shown in Fig.~\ref{fig:cond2Db}. If the Fermi level lies in the magnetic gap, we expect a vanishing longitudinal response and a quantized Hall response. This is shown in Fig.~\ref{fig:cond2Da}(b) and (c) for the gapped network, and in Fig.~\ref{fig:cond2Db}(b) and (c) for weakly-coupled ZZ branches. Note that the Hofstadter pattern in the response matches well to the magnetic bands. However, as we consider a finite system, the Hall conductance $G_H$ is not perfectly quantized. Indeed, due to tunneling processes the longitudinal conductance does not vanish completely in the magnetic gaps. Such finite-size effects also limit the resolution of the Hofstadter pattern.

\begin{figure}
\centering
\includegraphics[width=\linewidth]{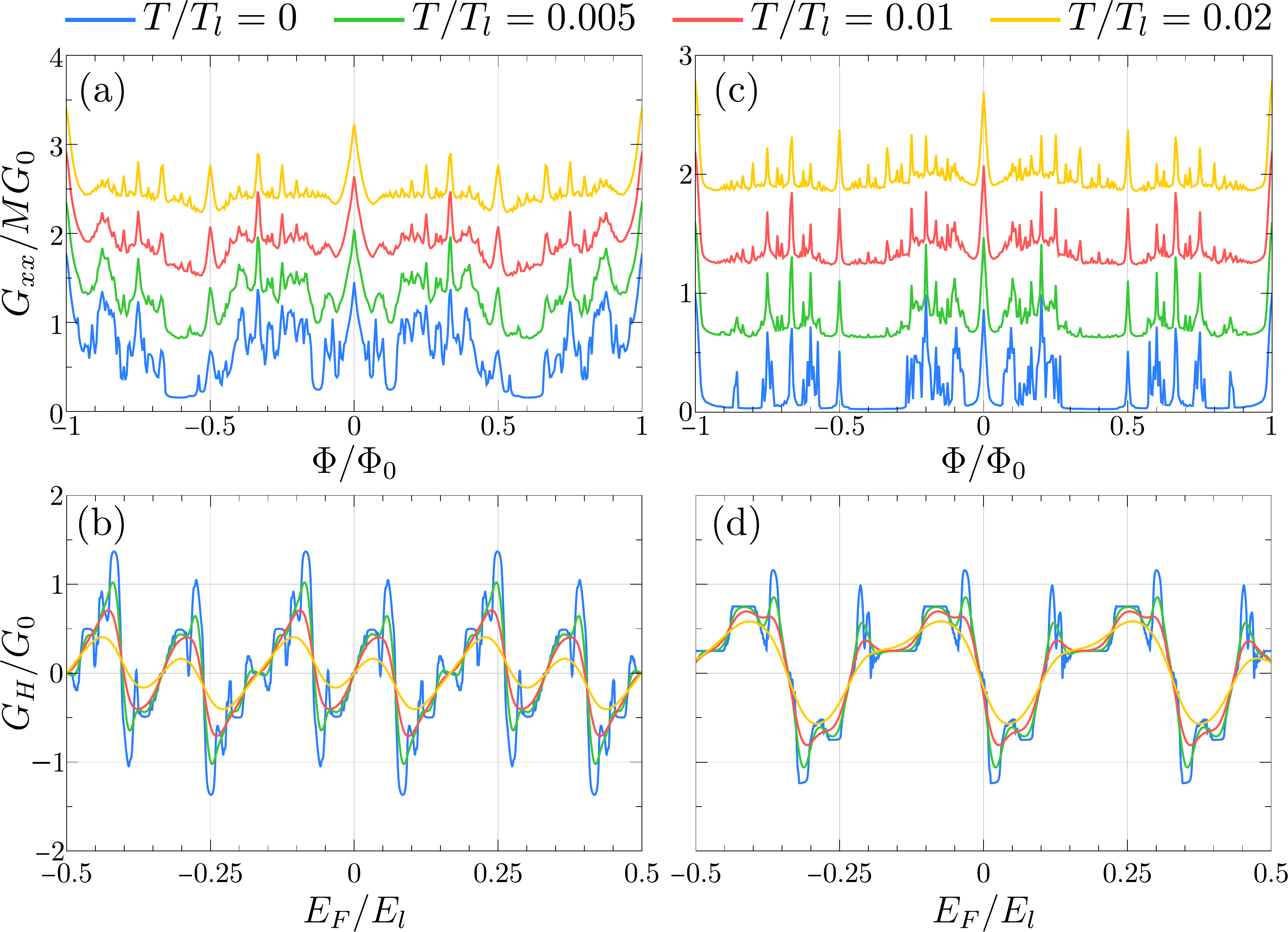}
\caption{Temperature dependence of the magnetoconductance. (a,c) Longitudinal response as a function of the magnetic flux $\Phi$ per moir\'e cell for $E_F / E_l = 0.1$ [horizontal dashed line in Fig.~\ref{fig:cond2Db}(b)] for the one and two channel network, respectively, with $T_l \approx 3400 \left( \theta^\circ \right)$~K. Network parameters correspond to Figs.\ \ref{fig:cond2Db} and \ref{fig:1channel}. For visibility, the curves are shifted by increments of $0.6$ with increasing temperature. (b,d) Hall response as a function of the Fermi energy for $B=4 \left( \theta^\circ \right)^{-2}$~T [vertical dashed line in Fig.~\ref{fig:cond2Db}(c)].}
\label{fig:temp}
\end{figure}

Finally, we address the temperature dependence of the conductance, which is shown in Fig.~\ref{fig:temp} for the same scattering parameters as in Fig.~\ref{fig:cond2Db}. As we mentioned before, the A-B oscillations due to chiral zigzag processes are due to interferences between paths that accumulate the same dynamical phase, while contributions from scattering between different zigzag branches do accumulate a relative phase. Hence the latter are suppressed at finite temperatures due to averaging over dynamical phases, which is demonstrated in Fig.~\ref{fig:temp}(a), where we introduced a network temperature scale $T_l = k_B / E_l \approx 3400 \left( \theta^\circ \right)$~K. On the other hand, the Hall response shown in Fig.~\ref{fig:temp}(b) is always suppressed with temperature since processes giving rise to the A-B oscillations do not contribute to the Hall response. 
At this point, we would like to point out some similarities to mesoscopic A-B rings. Similar to the network, one can find two kinds of interferences in A-B rings, namely those between paths accumulating the same or a different dynamical phase \cite{Stone1985,Webb1985, Nazarov}. The latter give rise to so-called universal conductance fluctuations which are sensitive to system parameters such as the electron density. In contrast, the former are due to time-reversed paths and their contributions survive ensemble averaging \cite{Umbach1986}. In the network, however, contributions independent of the dynamical phase do not correspond to interference between time-reversed paths because they occur within a single valley. In this respect, the A-B physics of the network is less robust since irregularities of the lattice or the electron density can locally alter the dynamical phase \cite{DeBeule2020a}.

\begin{figure*}
\centering
\includegraphics[width=\linewidth]{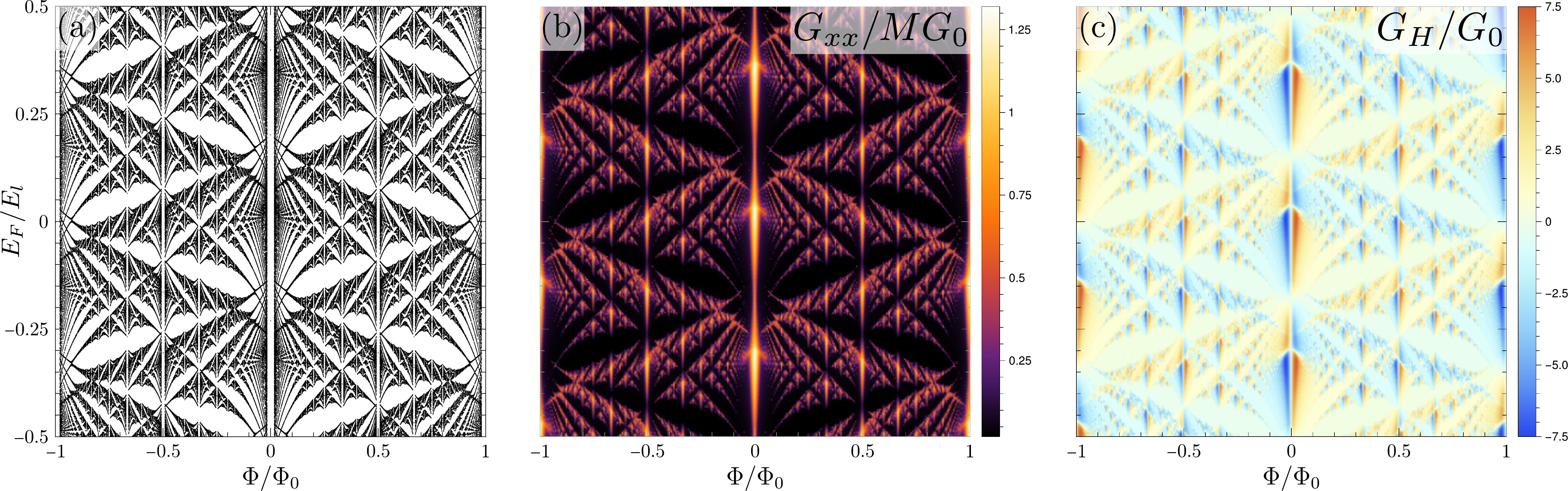}
\caption{(a) Hofstadter butterfly of the one-channel triangular network with $C_2T$ symmetry for $P_f = 0.7$. (b) Longitudinal and (c) Hall response for the four-terminal setup with $N=30$ and $M=60$, for the same parameters as in (a) with $G_0 = 4e^2/h$.}
\label{fig:1channel}
\end{figure*}

The one-channel model exhibits a similar phenomenology, as shown in Figs.\ \ref{fig:temp}(c) and (d) and Fig.~\ref{fig:1channel}. At temperatures $T \ll T_l$, all possible periods are present, but as the temperature is increased, only contributions independent of the dynamical phase survive. This gives rise to a change in periodicity from $2\Phi_0$ to $\Phi_0$ as the temperature is increased. There are also some differences in comparison with the two-channel case. Firstly, the oscillations in the longitudinal conductance do not appear on a constant background and the main periodicity at zero temperature always corresponds to two flux quanta per moir\'e cell, as opposed to one flux quantum in the chiral zigzag regime. Secondly, the energy bands obtained with the one-channel theory do not reproduce microscopic band structure calculations \cite{Fleischmann2020,Tsim2020}. In contrast, the two-channel  theory reproduces the energy dispersion and predicts robust A-B oscillations in the longitudinal response. Hence, we believe the observed A-B oscillations in the experiment of Ref.\ \onlinecite{Xu2019} can be understood in terms of scattering between parallel zigzag channels. Moreover, while the Hall response vanishes in the chiral zigzag regime, a finite Hall response is obtained when different zigzag branches are weakly coupled. At the same time, the A-B oscillations persist and tend to dominate at sufficiently high temperatures.

\section{Conclusions} \label{sec:conclusions}

We constructed a two-channel network model for the network of valley Hall states that emerges in minimally twisted bilayer graphene in the presence of an interlayer bias. To this end, we constrained the $S$ matrix of a single scattering node with the symmetries of the bilayer system. In the absence of forward scattering, we find that the model is characterized by a single scattering parameter, given by the relative phase shift that is acquired after deflections at a scattering node. This parameter tunes the network between pseudo-Landau levels and chiral zigzag modes, such that both phenomena are captured by a single network theory. We proceeded to include forward scattering and studied how additional scattering processes affect the chiral zigzag modes. We find that there are two regimes, depending on the couplings between zigzag modes. In the chiral zigzag regime, only parallel zigzag modes are coupled, such that the network effectively decouples into sets of quasi one-dimensional systems. On the other hand, scattering between zigzag modes that propagate in different directions gives rise to a two-dimensional percolating network.  

Subsequently, we used the network model to study electronic transport in the presence of a perpendicular magnetic field for a four-terminal setup. In the zigzag regime, we find that forward scattering gives rise to robust Aharonov-Bohm oscillations in the longitudinal conductance, while the Hall response vanishes. We also investigated the effect of a uniform in-plane electric field in this regime. Surprisingly, we find that the electric field leads to exactly the same oscillations at zero magnetic field. Moreover, when both the magnetic and electric field are present, the Aharonov-Bohm resonances of different chiral zigzag modes can be separated. Hence, this effect could be a further confirmation for the existence of chiral zigzag modes in the network. In order to obtain a finite Hall response, we introduced a weak coupling between different zigzag modes. In this case, the network supports Hofstadter physics at experimentally accessible magnetic fields. Furthermore, when the coupling between the different zigzag branches is not too strong, the Aharonov-Bohm resonances survive and are expected to dominate at finite temperatures.

To conclude, we demonstrated how one can construct a scattering model for the topological network that emerges in minimally twisted bilayer graphene under an interlayer bias, using the symmetries of the system. We then performed four-terminal transport calculations with this network model. Our results are consistent with previous transport experiments and suggest possible new experiments to further probe the nature of the topological network in minimally twisted bilayer graphene.

\begin{acknowledgments}
We thank R.\ F.\ Werner and Ming-Hao Liu for fruitful and interesting discussions. F.D.\ and P.R.\ gratefully acknowledge funding by the Deutsche Forschungsgemeinschaft (DFG, German Research Foundation) within the framework of Germany's Excellence Strategy -- EXC-2123 QuantumFrontiers -- 390837967.
\end{acknowledgments}

\appendix

\section{General scattering matrix} \label{app:model}

We first consider the one-channel case. We construct incoming modes in a single valley that are angular momentum $L_z$ eigenstates. These modes form an irreducible representations of the $C_3$ group,
\begin{equation}
a_m = \frac{1}{\sqrt{3}} \left( a_1 + \eta^{m} a_2 + \eta^{-m} a_3 \right),
\end{equation}
with $\eta = e^{-i2\pi/3}$ where $m = 0, \pm$ corresponds to the $L_z$ eigenvalue. They are thus eigenstates of $C_3$ ($a_1 \rightarrow a_2 \rightarrow a_3 \rightarrow a_1$) with $e^{-i \frac{2 \pi}{3} L_z / \hbar} a_m = \eta^m a_m$. Since the angular momentum is conserved, we have
\begin{equation}
\begin{pmatrix}
b_+ \\ b_0 \\ b_-
\end{pmatrix}
= \underbrace{
\begin{pmatrix} e^{i \xi_+} & 0 & 0 \\ 0 & e^{i \xi_0} & 0 \\ 0 & 0 & e^{i \xi_-} \end{pmatrix}}_{S}
\begin{pmatrix}
a_+ \\ a_0 \\ a_-
\end{pmatrix},
\end{equation}
with $\xi_m$ real. In the original basis, 
$\mathcal S = V^\dag S V$ with
\begin{equation}
V = \frac{1}{\sqrt{3}} \begin{pmatrix} 1 & \eta^* & \eta \\ 1 & 1 & 1 \\ 1 & \eta & \eta^* \end{pmatrix}.
\end{equation}

If in addition, $C_2$ symmetry is present, the eigenstates of the full $S$ matrix, including both valleys, are given by
\begin{equation}
a_{m}^{(\pm)} = \frac{1}{\sqrt{2}} \left( a_m \pm a_m' \right),
\end{equation}
where the prime indicates the other valley and which transform properly under both $C_2$ and $C_3$. In this basis, we thus have
\begin{widetext}
\begin{equation}
\begin{pmatrix}
b_{+}^{(+)} \\ b_{0}^{(+)} \\ b_{-}^{(+)} \\ b_{+}^{(-)} \\ b_{0}^{(-)} \\ b_{-}^{(-)}
\end{pmatrix} = \begin{pmatrix} e^{i \xi_{+}^{(+)}} & 0 & 0  & 0 & 0 & 0 \\ 0 & e^{i \xi_{0}^{(+)}} & 0 & 0 & 0 & 0 \\ 0 & 0 & e^{i \xi_{-}^{(+)}} & 0 & 0 & 0 \\ 0 & 0 & 0 & e^{i \xi_{+}^{(-)}} & 0 & 0 \\ 0 & 0 & 0 & 0 & e^{i \xi_{0}^{(-)}} & 0 \\ 0 & 0 & 0 & 0 & 0 & e^{i \xi_{-}^{(-)}} \end{pmatrix}\begin{pmatrix}
a_{+}^{(+)} \\ a_{0}^{(+)} \\ a_{-}^{(+)} \\ a_{+}^{(-)} \\ a_{0}^{(-)} \\ a_{-}^{(-)}
\end{pmatrix},
\end{equation}
\end{widetext}
where the valleys are decoupled if both eigenstates acquire the same phase shift $\xi_{m}^{(+)} = \xi_{m}^{(-)}$. Indeed,
\begin{align}
b_m & = \frac{1}{\sqrt{2}} \left( b_{m}^{(+)} + b_{m}^{(-)} \right) \\
& = \frac{e^{i \xi_{m}^{(+)}} + e^{i \xi_{m}^{(-)}}}{2} \, a_m + \frac{e^{i \xi_{m}^{(+)}} - e^{i \xi_{m}^{(-)}}}{2} \, a_m'.
\end{align}
The $S$ matrix in the original basis is now obtained with the transformation
\begin{equation}
Q = \frac{1}{\sqrt{2}} \begin{pmatrix} V & +V \\ V & -V \end{pmatrix},
\end{equation}
giving $\mathcal S = \mathcal S'$.

Finally, we consider time-reversal symmetry. Time reversal flips the valley and $L_z$, and exchanges incoming and outgoing modes:
\begin{equation}
a_m \rightarrow \left( b_{-m}' \right)^*, \qquad b_m \rightarrow \left( a_{-m}' \right)^*.
\end{equation}
We find the following condition on the $S$ matrix in the angular momentum basis,
\begin{equation}
\underline{S} = P \underline{S}^t P,
\end{equation}
where the underline denotes the total $S$ matrix containing both valleys and
\begin{equation}
P = 
\begin{pmatrix}
0 & 0 & 1 & & & \\
0 & 1 & 0 & & & \\
1 & 0 & 0 & & & \\
& & & 0 & 0 & -1 \\
& & & 0 & -1 & 0 \\
& & & -1 & 0 & 0 \\
\end{pmatrix},
\end{equation}
with $P^2 = 1$ and which implies $\xi_{+}^{(\pm)} = \xi_{-}^{(\pm)}$. Moreover, in the original basis, we have
\begin{equation}
\underline{\mathcal S} = \left( Q^\dag P Q \right) \left[ \left( Q^t Q \right)^\dag \underline{\mathcal S}^t \left( Q^t Q \right) \right] \left( Q^\dag P Q \right),
\end{equation}
where $\left( Q^t Q \right)^\dag \underline{\mathcal S}^t \left( Q^t Q \right)  = \underline{\mathcal S}$. This yields,
\begin{equation}
s_f = s_f', \qquad s_r = s_l', \qquad s_l = s_r'.
\end{equation}

When both $C_2$ and $T$ are present, we conclude that the $S$ matrix for the single-channel network has one real parameter up to a global phase.

We generalize this approach to $k$ channels per link. The $S$ matrix in the $L_z$ basis now takes the form
\begin{equation}
S = U_+(k) \oplus U_0(k) \oplus U_-(k),
\end{equation}
where $U_m(k)$ is a unitary matrix of dimension $k$, giving $3k^2$ real parameters. The $U$ matrices describe scattering processes between channels with the same angular momentum. Similarly as before, the unitary matrices are equal for the even and odd $C_2$ states in the absence of intervalley coupling, while $T$  symmetry implies $U_+(k) = U_-(k)^t$ and $U_0(k) = U_0(k)^t$. If $C_2T$ is conserved, there are $\left[ k^2 - k(k-1)/2 \right] + k^2= k(3k+1)/2$ real parameters. For the network in mTBG ($k=2$) there are six parameters (not including a global phase) and an explicit expression can be obtained with
\begin{equation}
U(2) = e^{i\varphi/2} \begin{pmatrix} e^{i \phi_1} \cos \theta & e^{i \phi_2} \sin \theta \\ -e^{-i \phi_2} \sin \theta & e^{-i \phi_1} \cos \theta \end{pmatrix},
\end{equation}
where $\phi_2 = \pi/2$ for $U_0$.

\section{Magnetic network bands and Hofstadter butterfly} \label{app:hof}

\begin{figure}
\centering
\includegraphics[width=\linewidth]{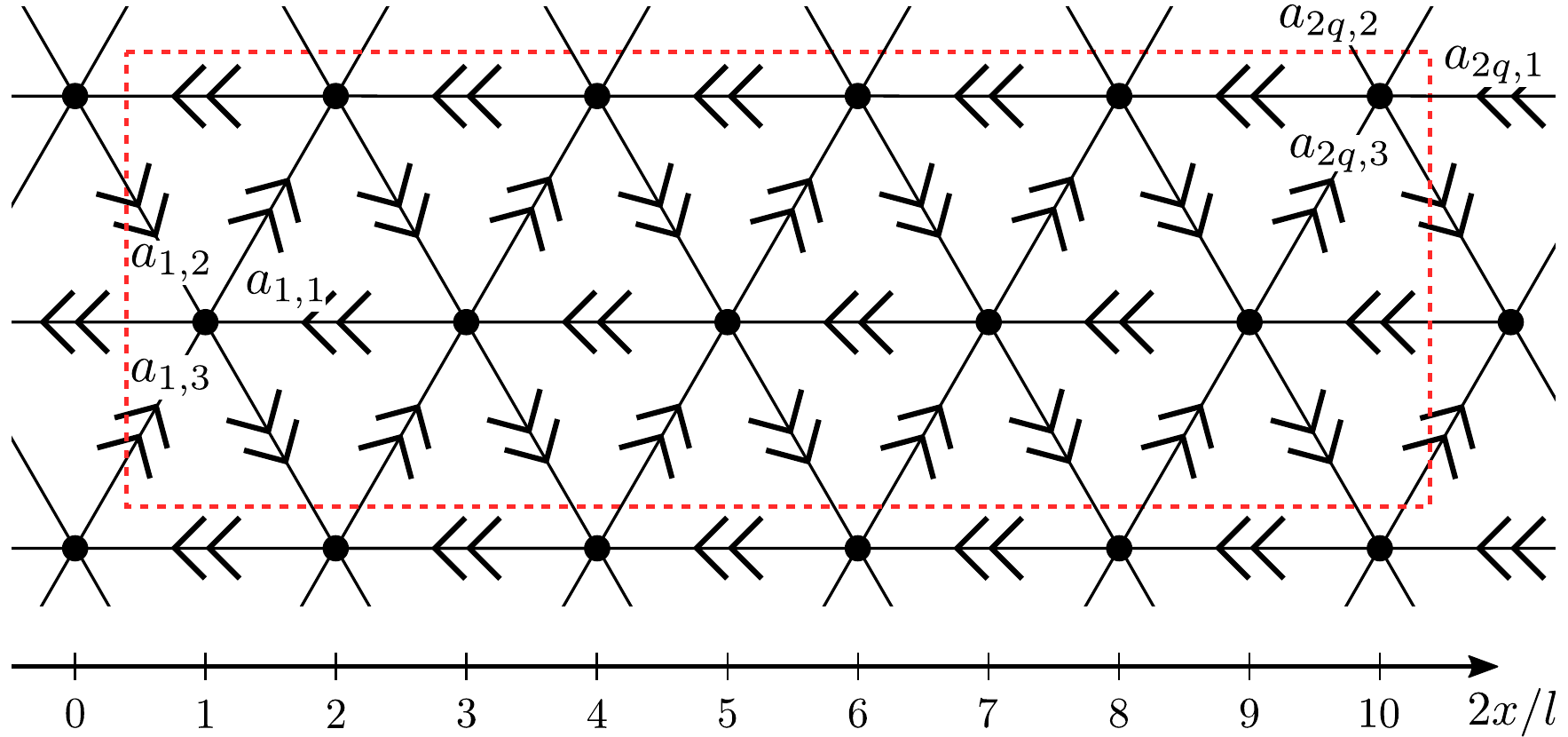}
\label{fig:mCell}
\caption{Magnetic unit cell (dashed rectangle) of the network in a perpendicular magnetic field for the Landau gauge $\bm A=B \bm e_y$, shown for a commensurate flux with $q=5$.}
\end{figure}
In the Landau gauge $\bm A = Bx \bm e_y$, the commensuration condition is obtained by demanding that the Peierls phase returns to itself after a translation $x \rightarrow x + ql$, where $q$ is a positive integer. We find
\begin{equation}
\frac{\pi \Phi}{\Phi_0} 2q = 2\pi p \quad \Rightarrow \quad \frac{\Phi}{\Phi_0} = \frac{p}{q},
\end{equation}
where $p$ is an integer and where all possible cases are obtained by taking $p$ and $q$ coprime. Thus, the magnetic unit cell has dimensions $ql \times \sqrt{3} l$, which is illustrated in Fig. \ref{fig:mCell} for $q=5$. If we label the amplitudes of incoming and outgoing modes in the magnetic cell as
\begin{align}
A_q & = \left( a_{1,1}, a_{1,1}', \ldots, a_{1,3}, a_{1,3}', \ldots, a_{2q,3}, a_{2q,3}' \right), \\
B_q & = \left( b_{1,1}, b_{1,1}', \ldots, b_{1,3}, b_{1,3}', \ldots, b_{2q,3}, b_{2q,3}' \right),
\end{align}
respectively, where the first index labels the scattering nodes and the second index labels the channels, as shown in Fig.~\ref{fig:lattice}(d). Here, we label the nodes by their horizontal position $x=nl/2$ with $n=1,\ldots,2q$. In this basis, the total $S$ matrix becomes
\begin{equation}
S_q = \mathds 1_{2q} \otimes \mathcal S,
\end{equation}
where $\mathcal S$ is the $S$ matrix for a single node, which is given in Eq.~\eqref{eq:C3}, and $B_q = S_q A_q$. Incoming and outgoing modes are also related by
\begin{equation}
e^{-i\varepsilon} A_q = M_q(\bm k,\Phi) B_q =  M_q(\bm k,\Phi) S_q A_q,
\end{equation}
where $M_q(\bm k,\Phi)$ is a $12q \times 12 q$ matrix that connects incoming and outgoing amplitudes of different nodes. For example, we have
\begin{align}
a_{1,1} & = e^{i\varepsilon} b_{3,1}, \\
a_{1,2} & = e^{i\varepsilon} e^{-ik_xql} b_{2q,2}, \\
a_{1,3} & = e^{i\varepsilon} e^{-ik_xql} e^{-ik_y \sqrt{3} l} b_{2q,3}, \\
a_{2,1} & = e^{i\varepsilon} b_{4,1}, \\
a_{2,2} & = e^{i\varepsilon} e^{ik_y \sqrt{3} l} e^{i\Phi_P(1)} b_{1,2}, \\
a_{2,3} & = e^{i\varepsilon} e^{-i\Phi_P(1)} b_{1,3},
\end{align}
and similar for the other channel. The magnetic network bands are then obtained by calculating the phase of the eigenvalues of the matrix $M_q(\bm k,\Phi) S_q$. The Hofstadter butterflies shown in Figs.\ \ref{fig:cond2Da}, \ref{fig:cond2Db}, and \ref{fig:1channel} are obtained by collecting the spectra for different fluxes at $\bm k=0$.

\section{Four-terminal setup} \label{app:4terminal}

Here, we give a detailed overview of the calculation of the total $S$ matrix for the four-terminal setup of the scattering network as shown in Fig.~\ref{fig:4terminal} with length $L=Nl$ and width $W=M\sqrt{3} l$, where $M,N=1,2,\ldots$. The incoming and outgoing modes at the first column of nodes in Fig.~\ref{fig:4terminal} are related by
\begin{equation}
\begin{pmatrix}
b_L^{(1)} \\ b_R^{(1)}
\end{pmatrix}
= \underbracket{\begin{pmatrix}
r_L^{(1)} & t_{LR}^{(1)} \\
t_{RL}^{(1)} & r_R^{(1)}
\end{pmatrix}}_{S_1~\textrm{in Fig.~\ref{fig:4terminal}}}
\begin{pmatrix}
a_L^{(1)} \\ a_R^{(1)}
\end{pmatrix},
\end{equation}
where $a_L^{(1)}$ ($a_R^{(1)}$) contain amplitudes of incoming modes at the left-hand (right-hand) side of the first section as shown in Fig.~\ref{fig:4terminal} and similar for outgoing modes. Here, the $S$ matrix for the first column is given by
\begin{equation}
S_1 =
\left(
\begin{array}{ccc|cccc}
r & & & t' & & & \\
& \ddots & & & \ddots & & \\
& & r & & & t' & \\
& & & & & & \mathds 1_{N_c} \\ \hline
t & & & r' & & & \\
& \ddots & & & \ddots & & \\
& & t & & & r' &
\end{array}
\right),
\end{equation}
where any omitted entries correspond to zeros and which has dimension $(4M+1)N_c$ where $N_c$ is the number of chiral channels per link. We have also defined
\begin{alignat}{2}
& r = \begin{pmatrix} 0 & 0 \\ s_r & s_l \end{pmatrix}, \qquad
&& t = \begin{pmatrix} s_l & s_f \\ s_f & s_r \end{pmatrix}, \\[2mm]
& r' = \begin{pmatrix} 0 & s_r \\ 0 & s_l \end{pmatrix}, \qquad
&& t' = \begin{pmatrix} 1 & 0 \\ 0 & s_f \end{pmatrix}.
\end{alignat}
The second column in Fig.~\ref{fig:4terminal} also contains modes of the up and down leads. Hence, we can write
\begin{equation}
\begin{pmatrix}
b_L^{(2)} \\ b_R^{(2)} \\ b_U^{(2)} \\ b_D^{(2)}
\end{pmatrix}
= \underbracket{\begin{pmatrix}
r_L^{(2)} & t_{LR}^{(2)} & t_{LU}^{(2)} & t_{LD}^{(2)} \\
t_{RL}^{(2)} & r_R^{(2)} & t_{RU}^{(2)} & t_{RD}^{(2)} \\
t_{UL}^{(2)} & t_{UR}^{(2)} & r_U^{(2)} & t_{UD}^{(2)} \\
t_{DL}^{(2)} & t_{DR}^{(2)} & t_{DU}^{(2)} & r_D^{(2)}
\end{pmatrix}}_{S_2~\textrm{in Fig.~\ref{fig:4terminal}}}
\begin{pmatrix}
a_L^{(2)} \\ a_R^{(2)} \\ a_U^{(2)} \\ a_D^{(2)}
\end{pmatrix},
\end{equation}
with
\begin{equation}
S_2 =
\left(
\begin{array}{ccccc|cccccc|c|c}
s_l & & & & & s_f & & & & & & s_r & \\
& r & & & & & t' & & & & & & \\
& & \ddots & & & & & \ddots & & & & & \\
& & & r & & & & & t' & & & & \\
& & & & & & & & & \mathds 1_{N_c} & & & \\
& & & & s_r & & & & & & s_f & & s_l \\ \hline
s_r & & & & & s_l & & & & & & s_f & \\
& t & & & & & r' & & & & & & \\
& & \ddots & & & & & \ddots & & & & & \\
& & & t & & & & & r' & & & & \\
& & & & s_l & & & & & & s_r & & s_f \\ \hline
s_f & & & & & s_r & & & & & & s_l & \\ \hline
& & & & s_f & & & & & & s_l & & s_r
\end{array}
\right),
\end{equation}
where again omitted entries are zero and which has dimension $(4M+3)N_c$. Incoming and outgoing modes of different columns are related by 
\begin{equation}
a_R^{(n)} = \alpha_n b_L^{(n+1)}, \qquad a_L^{(n+1)} = \beta_n b_R^{(n)}, \label{eq:combo}
\end{equation}
where for the clean network
\begin{equation}
\alpha_n = e^{i(\varepsilon-V_n)/2}, \quad \beta_n = e^{i(\varepsilon-V_n)} A((-1)^{n+1}n\Phi),
\end{equation}
with $\varepsilon=El/\hbar v$ the dynamical phase, $V_n=2\pi V(x_n)/E_l = 2\pi n V_0/E_l$ the value of the scalar potential at $x_n = ( l/2 ) ( n + 1/2 )$ [Fig.~\ref{fig:fields}(b)], and $A(z) = \mathds 1_M \otimes \left[ \exp \left( -i z \pi \sigma_z /\Phi_0 \right) \otimes \mathds 1_{N_c} \right]$ gives the Peierls phase in the Landau gauge $\bm A = Bx \bm e_y$. In the presence of smooth charge disorder, the matrices $\alpha_n$ and $\beta_n$ will also contain random phases. 

The total $S$ matrix for the first and second column is denoted as $(S_1\times S_2)_{\textrm{I}}$. Hence, we can write
\begin{equation}  \label{eq:S12}
\begin{pmatrix}
b_L^{(1)} \\ b_R^{(2)} \\ b_U^{(2)} \\ b_D^{(2)}
\end{pmatrix}
= \underbracket{\begin{pmatrix}
r_L^{(\textrm{I})} & t_{LR}^{(\textrm{I})} & t_{LU}^{(\textrm{I})} & t_{LD}^{(\textrm{I})} \\
t_{RL}^{(\textrm{I})} & r_R^{(\textrm{I})} & t_{RU}^{(\textrm{I})} & t_{RD}^{(\textrm{I})} \\
t_{UL}^{(\textrm{I})} & t_{UR}^{(\textrm{I})} & r_U^{(\textrm{I})} & t_{UD}^{(\textrm{I})} \\
t_{DL}^{(\textrm{I})} & t_{DR}^{(\textrm{I})} & t_{DU}^{(\textrm{I})} & r_D^{(\textrm{I})}
\end{pmatrix}}_{\left( S_1 \times S_2 \right)_{\textrm{I}}}
\begin{pmatrix}
a_L^{(1)} \\ a_R^{(2)} \\ a_U^{(2)} \\ a_D^{(2)}
\end{pmatrix},
\end{equation}
where we eliminated $a_R^{(1)}$, $a_L^{(2)}$, $b_R^{(1)}$, and $b_L^{(2)}$, giving
\allowdisplaybreaks
\begin{align}
r_L^{(\textrm{I})} & = r_L^{(1)} + t_{LR}^{(1)} \alpha_1 Q_2^{(\textrm{I})} r_L^{(2)} \beta_1 t_{RL}^{(1)}, \label{eq:S12a} \\
t_{LR}^{(\textrm{I})} & = t_{LR}^{(1)} \alpha_1 Q_2^{(\textrm{I})} t_{LR}^{(2)}, \\
t_{LU}^{(\textrm{I})} & = t_{LR}^{(1)} \alpha_1 Q_2^{(\textrm{I})} t_{LU}^{(2)}, \\
t_{LD}^{(\textrm{I})} & = t_{LR}^{(1)} \alpha_1 Q_2^{(\textrm{I})} t_{LD}^{(2)}, \\
t_{RL}^{(\textrm{I})} & = t_{RL}^{(2)} \beta_1 Q_1^{(\textrm{I})} t_{RL}^{(1)}, \\
r_R^{(\textrm{I})} & = r_R^{(2)} + t_{RL}^{(2)} \beta_1 Q_1^{(\textrm{I})} r_R^{(1)} \alpha_1 t_{LR}^{(2)}, \\
t_{RU}^{(\textrm{I})} & = t_{RU}^{(2)} + t_{RL}^{(2)} \beta_1 Q_1^{(\textrm{I})} r_R^{(1)} \alpha_1 t_{LU}^{(2)}, \\
t_{RD}^{(\textrm{I})} & = t_{RD}^{(2)} + t_{RL}^{(2)} \beta_1 Q_1^{(\textrm{I})} r_R^{(1)} \alpha_1 t_{LD}^{(2)}, \\
t_{UL}^{(\textrm{I})} & = t_{UL}^{(2)} \beta_1 Q_1^{(\textrm{I})} t_{RL}^{(1)}, \\
t_{UR}^{(\textrm{I})} & = t_{UR}^{(2)} + t_{UL}^{(2)} \beta_1 Q_1^{(\textrm{I})} r_R^{(1)} \alpha_1 t_{LR}^{(2)}, \\
r_U^{(\textrm{I})} & = r_U^{(2)} + t_{UL}^{(2)} \beta_1 Q_1^{(\textrm{I})} r_R^{(1)} \alpha_1 t_{LU}^{(2)}, \\
t_{UD}^{(\textrm{I})} & = t_{UD}^{(2)} + t_{UL}^{(2)} \beta_1 Q_1^{(\textrm{I})} r_R^{(1)} \alpha_1 t_{LD}^{(2)}, \\
t_{DL}^{(\textrm{I})} & = t_{DL}^{(2)} \beta_1 Q_1^{(\textrm{I})} t_{RL}^{(1)}, \\
t_{DR}^{(\textrm{I})} & = t_{DR}^{(2)} + t_{DL}^{(2)} \beta_1 Q_1^{(\textrm{I})} r_R^{(1)} \alpha_1 t_{LR}^{(2)}, \\
t_{DU}^{(\textrm{I})} & = t_{DU}^{(2)} + t_{DL}^{(2)} \beta_1 Q_1^{(\textrm{I})} r_R^{(1)} \alpha_1 t_{LU}^{(2)}, \\
r_D^{(\textrm{I})} & = r_D^{(2)} + t_{DL}^{(2)} \beta_1 Q_1^{(\textrm{I})} r_R^{(1)} \alpha_1 t_{LD}^{(2)}, \label{eq:S12b}
\end{align}
with
\begin{align}
Q_1^{(\textrm{I})} & = \left[ \mathds 1_{2MN_c} - r_R^{(1)} \alpha_1 r_L^{(2)} \beta_1 \right]^{-1}, \\
Q_2^{(\textrm{I})} & = \left[ \mathds 1_{(2M+1)N_c} - r_L^{(2)} \beta_1 r_R^{(1)} \alpha_1 \right]^{-1}.
\end{align}
Assuming identical scattering nodes throughout the sample, we obtain the combined $S$ matrix for the $(2j-1)$th and $(2j)$th column in a similar way where $j=1,\ldots,N$. For example, for the third and fourth column, we have
\begin{equation} \label{eq:S34}
\begin{pmatrix}
b_L^{(3)} \\ b_R^{(4)} \\ b_U^{(4)} \\ b_D^{(4)}
\end{pmatrix}
= 
\underbracket{\begin{pmatrix}
r_L^{(\textrm{II})} & t_{LR}^{(\textrm{II})} & t_{LU}^{(\textrm{II})} & t_{LD}^{(\textrm{II})} \\
t_{RL}^{(\textrm{II})} & r_R^{(\textrm{II})} & t_{RU}^{(\textrm{II})} & t_{RD}^{(\textrm{II})} \\
t_{UL}^{(\textrm{II})} & t_{UR}^{(\textrm{II})} & r_U^{(\textrm{II})} & t_{UD}^{(\textrm{II})} \\
t_{DL}^{(\textrm{II})} & t_{DR}^{(\textrm{II})} & t_{DU}^{(\textrm{II})} & r_D^{(\textrm{II})}
\end{pmatrix}}_{\left( S_1 \times S_2 \right)_{\textrm{II}}}
\begin{pmatrix}
a_L^{(3)} \\ a_R^{(4)} \\ a_U^{(4)} \\ a_D^{(4)}
\end{pmatrix}
\end{equation}
where the scattering matrix in Eq.~\eqref{eq:S34} contains the same expressions \eqref{eq:S12a}-\eqref{eq:S12b} with phase matrices $\alpha_1, \beta_1 \rightarrow \alpha_3, \beta_3$. 
If we now combine the two two-column $S$ matrices in Eqs.\ \eqref{eq:S12} and \eqref{eq:S34}, we obtain
\begin{widetext}
\begin{equation} \label{eq:S14}
\begin{pmatrix}
b_L^{(1)} \\ b_R^{(4)} \\ b_U^{(2)} \\ b_U^{(4)} \\ b_D^{(2)} \\ b_D^{(4)}
\end{pmatrix}
= 
\begin{pmatrix}
r_L & t_{LR} & t_{LU,1} & t_{LU,2} & t_{LD,1} & t_{LD,1} \\
t_{RL} & r_R & t_{RU,1} & t_{RU,2} & t_{RD,1} & t_{RD,1} \\
t_{UL,1} & t_{UR,1} & r_{U,11} & r_{U,12} & t_{UD,11} & t_{UD,12} \\
t_{UL,2} & t_{UR,2} & r_{U,21} & r_{U,22} & t_{UD,21} & t_{UD,22} \\
t_{DL,1} & t_{DR,1} & t_{DU,11} & t_{DU,12}  & r_{D,11} & r_{D,12} \\
t_{DL,2} & t_{DR,2} & t_{DU,21} & t_{DU,22} & r_{D,21} & r_{D,22}
\end{pmatrix}
\begin{pmatrix}
a_L^{(1)} \\ a_R^{(4)} \\ a_U^{(2)} \\ a_U^{(4)} \\ a_D^{(2)} \\ a_D^{(4)}
\end{pmatrix}
\end{equation}
\end{widetext}
where the total $S$ matrix for the two columns now has dimension $(4M+5)N_c$. Here, we used
\begin{equation}
a_R^{(2)} = \alpha_2 b_L^{(3)}, \qquad a_L^{(3)} = \beta_2 b_R^{(2)},
\end{equation}
to obtain Eq.~\eqref{eq:S14} where
\begin{align}
r_L & = r_L^{(\textrm{I})} + t_{LR}^{(\textrm{I})} \alpha_2 Q_2 r_L^{(\textrm{II})} \beta_2 t_{RL}^{(\textrm{I})}, \\
t_{LR} & = t_{LR}^{(\textrm{I})} \alpha_2 Q_2 t_{LR}^{(\textrm{II})}, \\
t_{LU,1} & = t_{LU}^{(\textrm{I})} + t_{LR}^{(\textrm{I})} \alpha_2 Q_2 r_L^{(\textrm{II})} \beta_2 t_{RU}^{(\textrm{I})}, \\
t_{LU,2} & = t_{LR}^{(\textrm{I})} \alpha_2 Q_2 t_{LU}^{(\textrm{II})}, \\
t_{LD,1} & = t_{LD}^{(\textrm{I})} + t_{LR}^{(\textrm{I})} \alpha_2 Q_2 r_L^{(\textrm{II})} \beta_2 t_{RD}^{(\textrm{I})}, \\
t_{LD,2} & = t_{LR}^{(\textrm{I})} \alpha_2 Q_2 t_{LD}^{(\textrm{II})}, \\
t_{RL} & = t_{RL}^{(\textrm{II})} \beta_2 Q_1 t_{RL}^{(\textrm{I})}, \\
r_R & = r_R^{(\textrm{II})} + t_{RL}^{(\textrm{II})} \beta_2 Q_1 r_R^{(\textrm{I})} \alpha_2 t_{LR}^{(\textrm{II})}, \\
t_{RU,1} & = t_{RL}^{(\textrm{II})} \beta_2 Q_1 t_{RU}^{(\textrm{I})}, \\
t_{RU,2} & = t_{RU}^{(\textrm{II})} + t_{RL}^{(\textrm{II})} \beta_2 Q_1 r_R^{(\textrm{I})} \alpha_2 t_{LU}^{(\textrm{II})}, \\
t_{RD,1} & = t_{RL}^{(\textrm{II})} \beta_2 Q_1 t_{RD}^{(\textrm{I})}, \\
t_{RD,2} & = t_{RD}^{(\textrm{II})} + t_{RL}^{(\textrm{II})} \beta_2 Q_1 r_R^{(\textrm{I})} \alpha_2 t_{LD}^{(\textrm{II})}, \\
t_{UL,1} & = t_{UL}^{(\textrm{I})} + t_{UR}^{(\textrm{I})} \alpha_2 Q_2 r_L^{(\textrm{II})} \beta_2 t_{RL}^{(\textrm{I})}, \\
t_{UL,2} & = t_{UL}^{(\textrm{II})} \beta_2 Q_1 t_{RL}^{(\textrm{I})}, \\
t_{UR,1} & = t_{UR}^{(\textrm{I})} \alpha_2 Q_2 t_{LR}^{(\textrm{II})}, \\
t_{UR,2} & = t_{UR}^{(\textrm{II})} + t_{UL}^{(\textrm{II})} \beta_2 Q_1 r_R^{(\textrm{I})} \alpha_2 t_{LR}^{(\textrm{II})}, \\
r_{U,11} & = r_U^{(\textrm{I})} + t_{UR}^{(\textrm{I})} \alpha_2 Q_2 r_L^{(\textrm{II})} \beta_2 t_{RU}^{(\textrm{I})}, \\
r_{U,12} & =  t_{UR}^{(\textrm{I})} \alpha_2 Q_2 t_{LU}^{(\textrm{II})}, \\
r_{U,21} & = t_{UL}^{(\textrm{II})} \beta_2 Q_1 t_{RU}^{(\textrm{I})}, \\
r_{U,22} & = r_U^{(\textrm{II})} + t_{UL}^{(\textrm{II})} \beta_2 Q_1 r_R^{(\textrm{I})} \alpha_2 t_{LU}^{(\textrm{II})}, \\
t_{UD,11} & = t_{UD}^{(\textrm{I})} + t_{UR}^{(\textrm{I})} \alpha_2 Q_2 r_L^{(\textrm{II})} \beta_2 t_{RD}^{(\textrm{I})}, \\
t_{UD,12} & = t_{UR}^{(\textrm{I})} \alpha_2 Q_2 t_{LD}^{(\textrm{II})}, \\
t_{UD,21} & = t_{UL}^{(\textrm{II})} \beta_2 Q_1 t_{RD}^{(\textrm{I})}, \\
t_{UD,22} & = t_{UD}^{(\textrm{II})} + t_{UL}^{(\textrm{II})} \beta_2 Q_1 r_R^{(\textrm{I})} \alpha_2 t_{LD}^{(\textrm{II})}, \\
t_{DL,1} & = t_{DL}^{(\textrm{I})} + t_{DR}^{(\textrm{I})} \alpha_2 Q_2 r_L^{(\textrm{II})} \beta_2 t_{RL}^{(\textrm{I})}, \\
t_{DL,2} & = t_{DL}^{(\textrm{II})} \beta_2 Q_1 t_{RL}^{(\textrm{I})}, \\
t_{DR,1} & = t_{DR}^{(\textrm{I})} \alpha_2 Q_2 t_{LR}^{(\textrm{II})}, \\
t_{DR,2} & = t_{DR}^{(\textrm{II})} + t_{DL}^{(\textrm{II})} \beta_2 Q_1 r_R^{(\textrm{I})} \alpha_2 t_{LR}^{(\textrm{II})}, \\
t_{DU,11} & = t_{DU}^{(\textrm{I})} + t_{DR}^{(\textrm{I})} \alpha_2 Q_2 r_L^{(\textrm{II})} \beta_2 t_{RU}^{(\textrm{I})}, \\
t_{DU,12} & = t_{DR}^{(\textrm{I})} \alpha_2 Q_2 t_{LU}^{(\textrm{II})}, \\
t_{DU,21} & = t_{DL}^{(\textrm{II})} \beta_2 Q_1 t_{RU}^{(\textrm{I})}, \\
t_{DU,22} & = t_{DU}^{(\textrm{II})} + t_{DL}^{(\textrm{II})} \beta_2 Q_1 r_R^{(\textrm{I})} \alpha_2 t_{LU}^{(\textrm{II})}, \\
r_{D,11} & = r_D^{(\textrm{I})} + t_{DR}^{(\textrm{I})} \alpha_2 Q_2 r_L^{(\textrm{II})} \beta_2 t_{RD}^{(\textrm{I})}, \\
r_{D,12} & = t_{DR}^{(\textrm{I})} \alpha_2 Q_2 t_{LD}^{(\textrm{II})}, \\
r_{D,21} & = t_{DL}^{(\textrm{II})} \beta_2 Q_1 t_{RD}^{(\textrm{I})}, \\
r_{D,22} & = r_D^{(\textrm{II})} + t_{DL}^{(\textrm{II})} \beta_2 Q_1 r_R^{(\textrm{I})} \alpha_2 t_{LD}^{(\textrm{II})},
\end{align}
with
\begin{align}
Q_1 & = \left[ \mathds 1_{2MN_c} - r_R^{(\textrm{I})} \alpha_2 r_L^{(\textrm{II})} \beta_2 \right]^{-1}, \\
Q_2 & = \left[ \mathds 1_{(2M+1)N_c} - r_L^{(\textrm{II})} \beta_2 r_R^{(\textrm{I})} \alpha_2 \right]^{-1}.
\end{align}

The $S$ matrix of the total system can therefore be obtained by first combining pairs of sections given by $S_1$ and $S_2$. 
Then we combine all the two-column $S$ matrices with Eqs.\ \eqref{eq:combo} for even $n$, as we have illustrated for $n=2$.
Note that in each step, the $S$ matrix grows as two more incoming and outgoing modes are added from the up and down leads. In the last step, we add one more section given by $S_1$ where we use
\begin{equation}
a_R^{(2N)} = \alpha_{2N} b_L^{(2N+1)}, \qquad a_L^{(2N+1)} = \beta_{2N} b_R^{(2N)},
\end{equation}
such that the total $S$ matrix of the four-terminal system has dimension $(4M+2N+1)N_c$. The transmission functions then become
\begin{align}
T_{\alpha\beta}(\Phi) & = \textrm{Tr} \big[ t_{\alpha\beta}^\dag t_{\alpha\beta}  \big]_K + \textrm{Tr} \big[ t_{\alpha\beta}^\dag t_{\alpha\beta} \big]_{K'} \\
& = T_{\alpha \beta}^{(K)}(\Phi) + T_{\beta \alpha}^{(K)}(-\Phi) = T_{\beta\alpha}(-\Phi), \label{eq:Tab}
\end{align}
with $r_\alpha = t_{\alpha\alpha}$ and where $\alpha,\beta=L,R,U,D$ label the leads. Here, the transmission of valley $K'$ follows from $S_{K'}(\Phi) = \left[ S_K(-\Phi) \right]^t$. The current at each lead is given in linear response by \cite{Buttiker1986,Buttiker1988}
\begin{equation} \label{eq:currentApp}
I_\alpha = \sum_{\beta} G_{\alpha\beta} \left( V_\alpha - V_\beta \right) 
\end{equation}
with $\sum_\alpha I_\alpha = 0$, $V_\alpha$ is the voltage at lead $\alpha$, and
\begin{equation} \label{eq:GTApp} 
G_{\alpha\beta} = \frac{2e^2}{h} \int dE \, T_{\alpha\beta} \left( - \frac{\partial f_0}{\partial E} \right),
\end{equation}
is the conductance  between leads $\beta$ and $\alpha$ with $f_0$ the Fermi-Dirac distribution. Since the total current has to vanish, Eqs.\ \eqref{eq:currentApp} are overdetermined. Therefore, we take lead $R$ as the reference, for example, and obtain
\begin{widetext}
\begin{equation}
\begin{pmatrix} I_L \\ I_U \\ I_D \end{pmatrix} =
\begin{pmatrix}
G_{LR}+G_{LU}+G_{LD} & -G_{LU} & -G_{LD} \\
-G_{UL} & G_{UL} + G_{UR} + G_{UD} & -G_{UD} \\
-G_{DL} & -G_{DU} & G_{DL} + G_{DR} + G_{DU}
\end{pmatrix}
\begin{pmatrix} V_L \\ V_U \\ V_D \end{pmatrix}.
\end{equation}
If we now consider the case where $I_L = -I_R$ and $I_U = -I_D$, one finds \cite{Buttiker1986}
\begin{align}
I_L & = G_{xx} \left( V_L - V_R \right)  + G_{xy} \left( V_U - V_D \right), \\
I_U & = G_{yx} \left( V_L - V_R \right)  + G_{yy} \left( V_U - V_D \right),
\end{align}
where we defined the conductances
\interdisplaylinepenalty=10000
\begin{align}
G_{xx} & = G_{RL} + \frac{\left( G_{DL} + G_{UL} \right) \left( G_{RD} + G_{RU} \right)}{\Sigma}, \label{eq:Gxx} \\
G_{yy} & = G_{UD} + \frac{\left( G_{LD} + G_{RD} \right) \left( G_{UL} + G_{UR} \right)}{\Sigma}, \label{eq:Gyy} \\
G_{xy} & = \frac{G_{LD} G_{RU} - G_{LU} G_{RD}}{\Sigma}, \\ 
G_{yx} & = \frac{G_{DL} G_{UR} - G_{UL} G_{DR}}{\Sigma},     
\end{align}
with
\begin{align}
\Sigma(\Phi) & = G_{LD} + G_{LU} + G_{RD} + G_{RU} \\
& = G_{DL} + G_{UL} + G_{DR} + G_{UR} = \Sigma(-\Phi),
\end{align}
since $\sum_\beta G_{\beta \alpha} = \sum_\beta G_{\alpha \beta}$ and $G_{\alpha\beta}(\Phi) = G_{\beta\alpha}(-\Phi)$ ($\alpha, \beta = L ,R, U, D$). 
\end{widetext}
Hence, we see that the Onsager relations $G_{ij}(\Phi) = G_{ji}(-\Phi)$ ($i,j=x,y$) hold only if we sum up the transmission of both valleys (Eq.~\eqref{eq:Tab}). Note that we have written the longitudinal conductances in Eqs.\ \eqref{eq:Gxx} and \eqref{eq:Gyy} as the sum of a direct contribution (i.e., the two-terminal part) and a contribution that can be interpreted as a weighted probability to transmit to the longitudinal lead via the transverse leads. As such, the longitudinal conductance is not given by the sum of the conductance of each valley separately, since it also contains incoherent scattering between the valleys via the leads. In such processes, one first transmits in one valley to a transverse lead followed by a transmission to the longitudinal lead in the other valley. Finally, we obtain the Hall response
\begin{equation}
G_H = \frac{G_{xy} - G_{yx}}{2}. 
\end{equation}
\begin{figure}
\centering
\includegraphics[width=\linewidth]{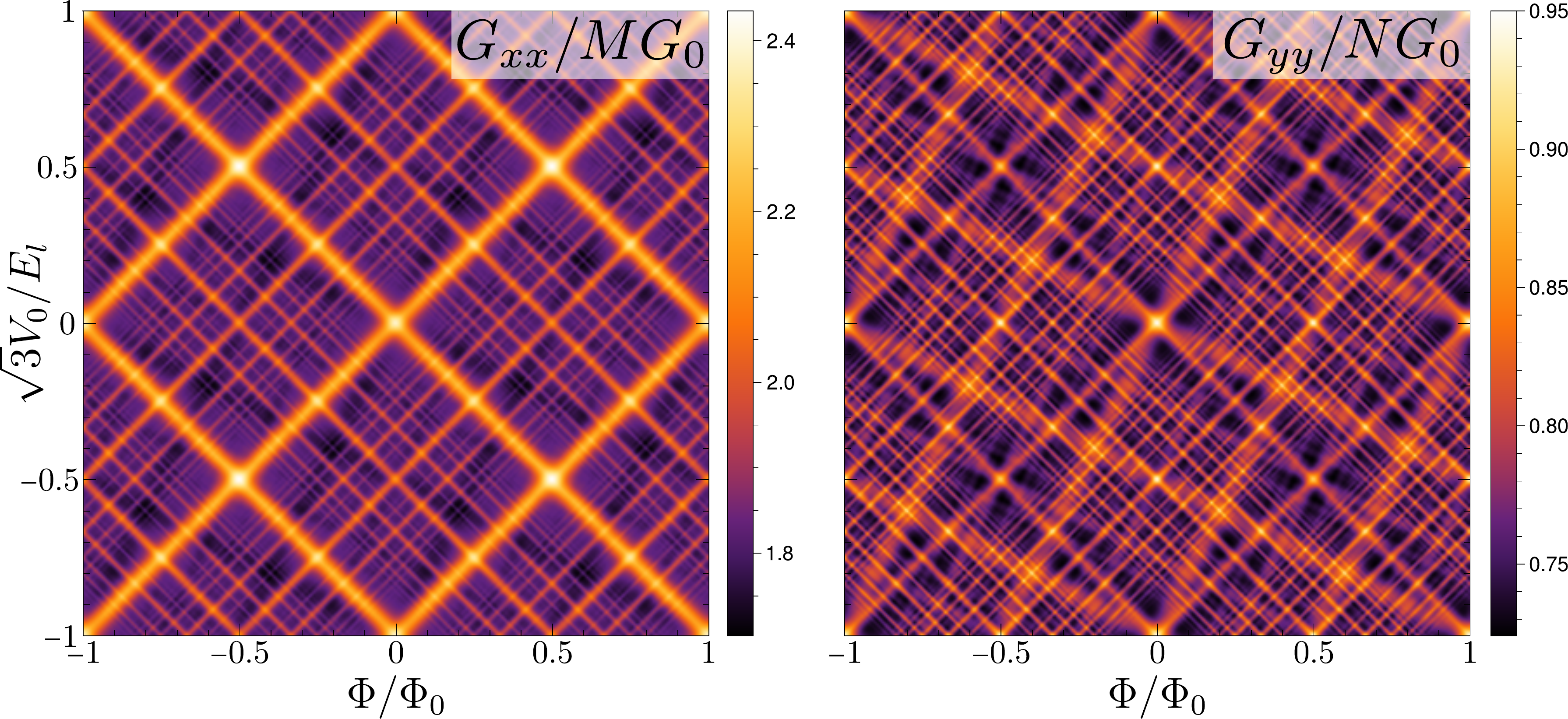}
\caption{Zero-temperature conductance $G_{xx}$ and $G_{yy}$ of the four-terminal setup as a function of the magnetic flux $\Phi$ per moir\'e cell and the slope of the potential $V(y) = 2V_0y / l$ for the same parameters as Fig.\ \ref{fig:condZZ}}
\label{fig:condEfieldY}
\end{figure}

\section{Dependence on the direction of the electric field}

To investigate the dependence on the direction of the in-plane electric field, we also performed four-terminal calculations for a constant electric field in the $y$ direction, as defined in Fig.\ \ref{fig:4terminal}. This is shown in Fig.\ \ref{fig:condEfieldY}. We find that the resonances in $G_{xx}$ are similar as for the electric field along the $x$ direction, except for the scale on the vertical axis. This is because the valley Hall channels along diagonal links accumulate a larger phase between nodes for the potential $V(y)$ as compared to $V(x)$ for the same electric field strength, since they span a larger distance in the $y$ direction. In contrast, the zigzag branch that propagates along the $x$ direction
is affected differently, which is clear from the square (anti)resonance pattern in $G_{yy}$ ($G_{xx}$)
as compared to the vertical lines in Fig.\ \ref{fig:condZZ}.

\bibliography{references}

\end{document}